\documentclass[aps,showpacs,nofootinbib,superscriptaddress]{revtex4}

\usepackage{graphicx}
\usepackage{dcolumn}
\usepackage{young}
\usepackage{wick}

\def\slashchar#1{\setbox0=\hbox{$#1$}
   \dimen0=\wd0 \setbox1=\hbox{/} \dimen1=\wd1
   \ifdim\dimen0>\dimen1 \rlap{\hbox to \dimen0{\hfil/\hfil}} #1
   \else  \rlap{\hbox to \dimen1{\hfil$#1$\hfil}} / \fi}
\begin{document}
\title{Large $N_c$ Weinberg-Tomozawa interaction and  
negative parity $s$-wave baryon resonances}
\author{C. Garc{\'\i}a-Recio}
\author{J. Nieves}
\author{L.L. Salcedo}
\affiliation{
Departamento de F{\'\i}sica 
At\'omica, Molecular y Nuclear, \\
Universidad de Granada, E-18071
Granada, Spain
}
\begin{abstract}
It is shown that in the 70 and 700 SU(6) irreducible spaces, the SU(6)
extension of the Weinberg-Tomozawa (WT) $s$-wave meson-baryon
interaction incorporating vector mesons ({\it hep-ph/0505233}) scales
as ${\cal O}(N_c^0)$, instead of the well known ${\cal O}(N_c^{-1})$
behavior for its SU(3) counterpart. However, the WT interaction
behaves as order ${\cal O}(N_c^{-1})$ within the 56 and 1134
meson-baryon spaces.  Explicit expressions for the WT couplings
(eigenvalues) in the irreducible SU(2$N_F$) spaces, for arbitrary
$N_F$ and $N_c$, are given.  This extended interaction is used as a
kernel of the Bethe-Salpeter equation, to study the large $N_c$
scaling of masses and widths of the lowest--lying negative parity
$s$-wave baryon resonances. Analytical expressions are found in the
$N_c\to \infty$ limit, from which it can be deduced that resonance
widths and excitation energies $(M_R-M)$ behave as order ${\cal O}
(N^0_c)$, in agreement with model independent arguments, and moreover
they fall in the 70-plet, as expected in constituent quark models for
an orbital excitation. For the 56 and 1134 spaces, excitation energies
and widths grow ${\cal O} (N_c^{1/2})$ indicating that such resonances
do not survive in the large $N_c$ limit. The relation of this latter
$N_c$ behavior with the existence of exotic components in these
resonances is discussed. The interaction comes out repulsive in the
700.

\end{abstract}


\maketitle



\section{Introduction}

 Quantum Chromodynamics (QCD), the theory of the strong interactions,
is a non-abelian gauge theory based on the gauge group SU($N_c$), with
the number of colors $N_c=3$. Several authors have pointed out that
many features of QCD can be understood by studying the $1/N_c$
expansion of the theory and that, even at the Leading Order (LO) $N_c
\to \infty$, non-trivial and realistic features can be
inferred~\cite{Ho74,Wi79,Ma99}.

The question of what is the true nature of baryon resonances has
attracted considerable attention in recent modern constructions of
effective field theories describing meson-baryon scattering. The
pattern of Spontaneous Chiral Symmetry Breaking (SCSB) of QCD, together
with an appropriate non-perturbative scheme, turns out to be a
crucial ingredient to better understand the main features of the
resonances. On the other hand, one might wonder what is the behavior
of these hadron states in the large $N_c$ limit of QCD.

To incorporate SCSB, we work in a recently developed framework to
describe meson-baryon, both in $s$- and $d$-waves (\cite{Ga03} and
\cite{KL03}, respectively), scattering and resonances.  It is based on
the solution of the Bethe Salpeter Equation (BSE) with a kernel
determined by the flavor SU(3) chiral counting rules and a particular
Renormalization Scheme (RS). The claim
of the authors of ~\cite{LK02} is that, in the SU(3) limit, this RS
restores crossing symmetry for a given total Center of Mass (CM)
energy ($\sqrt s$) below the unitarity threshold. At LO, in the chiral
expansion of the kernel, all parameters are determined, and the
obtained results are in a remarkable agreement with
data~\cite{Ga03, KL03,Lu03}. Extensions of the formalism to the
meson-meson sector~\cite{LK04} and the study of charm baryon
resonances~\cite{LK04charm} also lead to excellent results.

One of the findings of Ref.~\cite{Ga03} is the existence of two SU(3)
octets plus a singlet of $J^P=\frac12^-$ $s$-wave baryon resonances
($N(1535)$, $N(1650)$, $\Lambda(1405)$, $\Xi(1690)$, $\cdots$), which
are dynamically generated.  In this work, we aim at describing the
large $N_c-$dependence of their masses and widths.  It is well known
that in the $N_c \to \infty$ limit, the spin 3/2 baryon decuplet
($\Delta$, $\Sigma^*$, $\Xi^*$, $\Omega$) is degenerate to the nucleon
octet. Therefore, for consistency~\cite{DJM94}, such degrees of
freedom have to be considered, which will force us to work with a
larger spin-flavor symmetry group (SU(6)).  Spin-flavor symmetry in
the meson sector is not a direct consequence of large $N_c$. However,
vector mesons ($K^*, \rho,\omega, {\bar K}^{*}, \phi$) do exist, they
will couple to baryons and presumably will influence the properties of
the resonances.  Lacking better theoretical founded models to take
into account vector mesons, we study here the spin-flavor symmetric
scenario, as reasonable first step.

This paper is organized as follows. In the next section, we briefly
sketch the chiral unitary model of Ref.~\cite{Ga03}, and its LO $N_c$
limit is discussed in Sect.~\ref{sec:sec3}. The baryon decuplet and
vector meson nonet   effects
are considered in Sect.~\ref{sec:sec4}. First in
Subsect.~\ref{sec:sec4a} , we use the chiral Bethe Salpeter approach
to SU(6) meson-baryon scattering developed in Ref.~\cite{GNS05}. In
Subsect.~\ref{sec:nc}, we extend the latter model for arbitrary $N_c$
and present the final and more robust results of this work. Finally,
in Sect.~\ref{sec:sec5}, we present our main conclusions. There are
four appendices where some useful formulae of interest for
Sect.~\ref{sec:sec2} and Subsect.~\ref{sec:nc} are compiled.

\section{Chiral Bethe Salpeter Approach to  SU(3) Meson-Baryon
Scattering ($\chi-$BS(3))} 

\label{sec:sec2}

The leading term of the $s$-wave chiral meson-baryon Lagrangian is the well
known Weinberg-Tomozawa (WT) interaction~\cite{Wein-Tomo}. Since the
pioneering works of the group of Weise~\cite{weise}, and using the
WT Lagrangian as the input of the BSE\footnote{In some of the works,
the authors use the Lippmann-Schwinger equation instead of the
relativistic BSE.}, several approaches to $s$-wave baryon resonances in
different strangeness and isospin sectors have been carried
out~\cite{LK02,oset,OM,nrg}. From the theoretical point of view, the
used RS constitutes indeed the main difference among all of these
works (see Ref.~\cite{Lu03} for details). 

In order to find resonances in this approach, the coupled channel BSE
is solved, with an interaction kernel expanded in chiral perturbation
theory as formulated in~\cite{Pich95}. The involved hadrons are the
Goldstone pseudoscalar meson ($K,\pi,\eta,\bar K$) and the lowest
$\frac12^+$ baryon (${\rm N} ,\Sigma, \Lambda , \Xi$) octets.  The
solution for the coupled channel $s$-wave meson-baryon scattering
amplitude, $T(\sqrt{s})$ in the so called {\it on-shell}
scheme~\cite{LK02,EJmeson} where the offshellness of the BSE is
ignored, can be expressed in terms of a renormalized matrix of loop
functions, $J(\sqrt{s})$, and an effective on-shell interaction
kernel, $V(\sqrt{s})$, as follows\footnote{ The $T$ matrix defined in
Eq.~(\protect\ref{eq:scat-eq}), coincides with the $t$ matrix defined
in Eq.~(33) of the first entry of Ref.~\protect\cite{nrg}.}
\begin{eqnarray}
T(\sqrt{s}) =  \frac{1}{1- V(\sqrt{s})\,J(\sqrt{s})}\,V(\sqrt{s})
\,. \label{eq:scat-eq}
\end{eqnarray}
Thanks to the conservation of Isospin ($I$) and Strangeness ($S$), the
problem decouples into different $(I,S)$ sectors. In each sector,
there are several coupled channels, $N^{IS}$. For instance, in the
$(I,S)=(0,-1)$ sector $N^{IS}=4$ and the corresponding coupled channels
are $\pi \Sigma$ , $\eta \Lambda$, $\bar K N$ and $K \Xi$. Thus for a
given $(I,S)$ sector, all objects in Eq.~(\ref{eq:scat-eq}) are
square matrices of dimension $N^{IS}$ in the coupled channel
space.  The effective on-shell interaction kernel $V$ is
expanded in chiral perturbation theory. The chiral LO
interaction kernel $V(\sqrt{s})$, as determined by the WT interaction reads
\begin{eqnarray}
V^{IS}_{ab}(\sqrt{s}) =
D^{IS}_{ab} \frac{2\,\sqrt{s}-M_a-M_b}{4\,f^2} \,,
\label{eq:lowest}
\end{eqnarray}
where $M_a$ ($M_b$) is the baryon mass of the initial (final)
channel. The $D$'s matrices can be found in the
literature~\cite{oset,nrg} or deduced from Eq.~(\ref{eq:su6}) of
Subsect.~\ref{sec:sec4a} (see \cite{GNS05} for some more details).
The eigenvalues ($\lambda$'s) of the $D^{IS}$ matrices are 2,0,$-3,-3$
for both $IS=(1/2,0)$ and $IS =(1/2,-2)$, and 2,$-3,-3,-6$ and
2,0,0,$-3,-3$ for $IS=(0,-1)$ and $IS =(1,-1)$, respectively. Those
eigenvalues follow a pattern inferred from the SU(3) group
representation reduction
\begin{equation}
8\otimes8=27\oplus 10 \oplus 10^* \oplus 8_a \oplus 8_b \oplus 1 \label{eq:su3}
\end{equation}
being 
\begin{equation}
\lambda_{8_a}=\lambda_{8_b}\equiv\lambda_8=-3, \qquad \lambda_1=-6, \qquad 
\lambda_{10}=\lambda_{10^*}=0, \qquad \lambda_{27}=2, \label{eq:lambdas}
\end{equation}
the eigenvalues associated to octets, singlet, decuplet and antidecuplet,
and 27--plet SU(3) representations, respectively~\cite{Ga03}. 

On the other hand, the diagonal loop functions, $J^{IS}(\sqrt{s})$,
can be found in the Appendix~\ref{sec:app}.  The loop function
logarithmically diverges and one subtraction is needed to make it
finite. Such a freedom is fixed by the renormalization
condition~\cite{LK02}
\begin{eqnarray}
T^{IS}(\sqrt{s}= \mu) = V^{IS}(\mu ) \,, \qquad \mu = \mu (I,S) \label{eq:rsch}
\end{eqnarray}
with the choice
\begin{eqnarray}
\mu(1/2,-2)&=& m_\Xi, \quad \mu(0,-1)=m_\Lambda, 
\nonumber \\
 \mu(1,-1)& = &m_\Sigma,  \quad  \mu(1/2,0) = m_N
\label{eq:sub-choice}
 \end{eqnarray}
The renormalization condition
of Eq.~(\ref{eq:rsch}) is implemented by imposing that the
renormalized loop functions $J^{IS}_a(\sqrt{s}),~\forall a=1,\cdots ,
N^{IS}$, vanish at the appropriate points $\sqrt{s}=\mu(I,S)$. In this
way, all the constants ${\cal J}^{IS}_a ( s= (m_a + M_a)^2),
a=1,\cdots , N^{IS}$ in Eq.~(\ref{eq:defj0}) turn out to be completely
determined in terms of the involved baryon and meson
masses. Furthermore, taking the LO of the chiral expansion for the
interaction kernel, $V(\sqrt{s})$, as determined by the WT
interaction, there are no free parameters besides the meson ($m$'s)
and baryon ($M$'s) masses and the pion weak decay constant in the
chiral limit ($f\simeq 90$ MeV). At this chiral LO, the framework
leads already to excellent results for physical $s$-wave meson-baryon
scattering~\cite{Ga03} (an extension of the model to $d$-wave scattering
works also quite nicely~\cite{KL03}). Besides, the framework allows
also to study the dependence of the scattering process on the quark
masses, which made possible to unravel the SU(3) structure of the
lowest lying $s$-wave baryon resonances. The findings of
Ref.~\cite{Ga03} indicate that two full SU(3) octets plus an
additional singlet of $\frac12^-$ resonances are dynamically
generated. Some of them are the four stars $N(1535)$, $N(1650)$,
$\Lambda(1405)$, $\Lambda(1670)$ or the three stars $\Xi(1690)$
resonances. All these resonances appear in the sectors
$(I,S)=(\frac12,0),(0,-1),(1,-1)$ and $(\frac12,-2)$. Similar
conclusions, within a different RS, can be drawn from the work of
Ref.~\cite{Ji03}, though there only the strangeness $-1$ sector is
studied in detail.

\section{Large $N_c$ Limit of the $\chi-$BS(3) }

\label{sec:sec3}

The $N_c \to \infty$ limit of the LO $\chi-$BS(3) model is
particularly simple, since as discussed above, the model has no free
parameters besides the hadron masses and the pion weak decay constant
in the chiral limit. The $N_c \to \infty $ behavior of those
quantities  is well established (see f.i. Ref.~\cite{Ma99}), and
neglecting $1/N_c^\epsilon$  terms ($\epsilon > 0$), one finds 
\begin{eqnarray}
f(N_c) &\sim& f_0\times \sqrt{\frac{N_c}{3}}\\
M_a(N_c) &\sim&  M_0 \frac{N_c}{3} + b_1 \frac{\sqrt 3}{2}\left (
1 - \frac{3N_s^{(a)}}{N_c}\right)   \\
m_a (N_c) &\sim& m_a
\end{eqnarray}
with $M_0 \approx 1097$ MeV from the coefficient $a_0$ in Eq.~(7.4) of
Ref.~\cite{Ma99}, $b_1\approx -257$ MeV and $f_0\approx 90$ MeV. Note
that the number of strange quarks, $N_s$, could be a fraction of the
total number of quarks ($N_c$) of the colorless baryon.

Resonances manifest as poles in the fourth quadrant of the second
Riemann sheet of the $T-$matrix.  Positions of the poles,
\begin{equation}
s_R=M^2_R - i M_R \Gamma_R, 
\end{equation}
determine masses ($M_R$) and widths ($\Gamma_R$) of the resonances
while the residues for the different channels define the corresponding
branching ratios. As mentioned above, an exhaustive study of the
$J^P=\frac12^-$ $s$-wave baryon resonance properties for the $S=0,-1$
and $-2$ channels was performed in Ref.~\cite{Ga03}. In what follows,
we neglect the meson masses, which become truly massless Goldstone
bosons, and the $b_1$ term contribution to the baryon masses, since
they do not affect the LO $N_c \to \infty$ properties of those
resonances\footnote{The $N_c$ dependence of the correction
induced by finite meson masses can be
estimated by shifting the baryon mass by an amount of order $N_c^0$.} .  In
this way SU(3) flavor symmetry is also restored, and one has two
degenerate octets and one singlet of resonances. Indeed, for $N_c=3$
we essentially recover the ``light'' SU(3) limit introduced in
Ref.~\cite{Ga03}. Within this framework, our RS leads to the
conditions ${\cal J}^{IS}_a ( s= M^2)=0, a=1,\cdots , N^{IS}$, where
$M$ is the $N_c$ LO SU(3) baryon mass
\begin{equation}
M_a \sim M = M_0\times N_c/3,  \qquad \forall a
\end{equation}
For each $IS$ channel, the position of the poles is determined by 
\begin{equation}
\beta(s)\Big|_{s=s_R\equiv M^2_R-iM_R\Gamma_R} =  \lambda_i, \quad
i=1,8,10,10^*, 27
\label{eq:beta} 
\end{equation} 
with $M_R>M$ and $\Gamma_R>0$. Besides,  $\lambda_i$ are the 
eigenvalues of the real and symmetric matrix $D^{IS}$
(Eq.~(\ref{eq:lambdas}))  and the
dimensionless function $\beta(s)$   (see
Eqs.~(\ref{eq:scat-eq})--(\ref{eq:lowest})) reads
\begin{equation}
\beta(s) = \frac{2f^2}{J_{II}^{N_c \gg 1}(\sqrt s)(\sqrt s - M)}
\label{eq:beta_ant} 
\end{equation}
with $J_{II}^{N_c \gg 1}$ the loop function of Eq.~(\ref{eq:loopf})
with $M_a=M, m_a=0$, but defined in the second Riemann sheet. In the
fourth quadrant, it reads~\cite{nrg}
\begin{eqnarray}
  J_{II}^{N_c \gg 1}(\sqrt s) &=& 
\frac{(\sqrt s +M)^2}{2\sqrt s~(4\pi)^2}\left (\frac{s-M^2}{s}\right) 
\Big\{ \log |R(s)| + i {\rm Arg}(R(s))-3i\pi\Big\}
\label{eq:jota}
\end{eqnarray}
with $R(s)=(s-M^2)/M^2$ and ${\rm Arg}(R(s))$ should be taken in the
interval $[0,2\pi[$.  

The equation (\ref{eq:beta}) has solutions only for negative
eigenvalues, $\lambda_8$ and $\lambda_1$. Thus, at LO of the $N_c$
expansion only those $s$-wave $\frac 12^-$ resonant states ($N(1535)$,
$N(1650)$, $\Lambda(1405)$, $\Lambda(1670)$, $\Sigma(1620)$,
$\Xi(1620)$, $\Xi(1690)$, $\Lambda(1390)$\footnote{The $\Lambda(1390)$ state
corresponds to the SU(3) singlet
representation~\protect\cite{Ga03,nrg,Ji03,ORM05}. In this list of
resonances, there is a $\Sigma$ state missing. Perhaps, it could be
the $\Sigma(1750)$ resonance.}) belonging to the two octets and
singlet SU(3) representations are dynamically generated from Goldstone meson
($K,\pi,\eta,\bar K$) and the lowest $J^P=\frac12^+$ baryon (${\rm N}
,\Sigma, \Lambda , \Xi$ ) octets re-scattering. Reciprocally, LO $N_c$
results disfavor the existence of dynamically generated decuplet,
antidecuplet and 27-plet states. Though, this will change after the
inclusion of baryon decuplet and vector meson nonet effects in the
next section. For its nowadays interest, we
remark that LO $N_c$ $\chi-$BS(3) model  strongly disfavors that the $S=+1$
isoscalar $\Theta^+$ resonance, which would be the
isospin singlet state of the antidecuplet representation,
could be described just in terms of dynamical $KN$ resonant
re-scattering\footnote{Moreover, we
should  remind here that the WT chiral meson-baryon Lagrangian predicts a
vanishing on shell interaction kernel $V(\sqrt s)$, for isoscalar $KN$
scattering.}. Taking into account also $K^*$ and $\Delta$ degrees of
freedom, within a larger spin-flavor symmetry scheme, might
permit the existence of the {\it so called} pentaquarks~\cite{GNS05}.

\begin{figure}
\vspace{-2cm}
\centerline{\includegraphics[height=25cm]{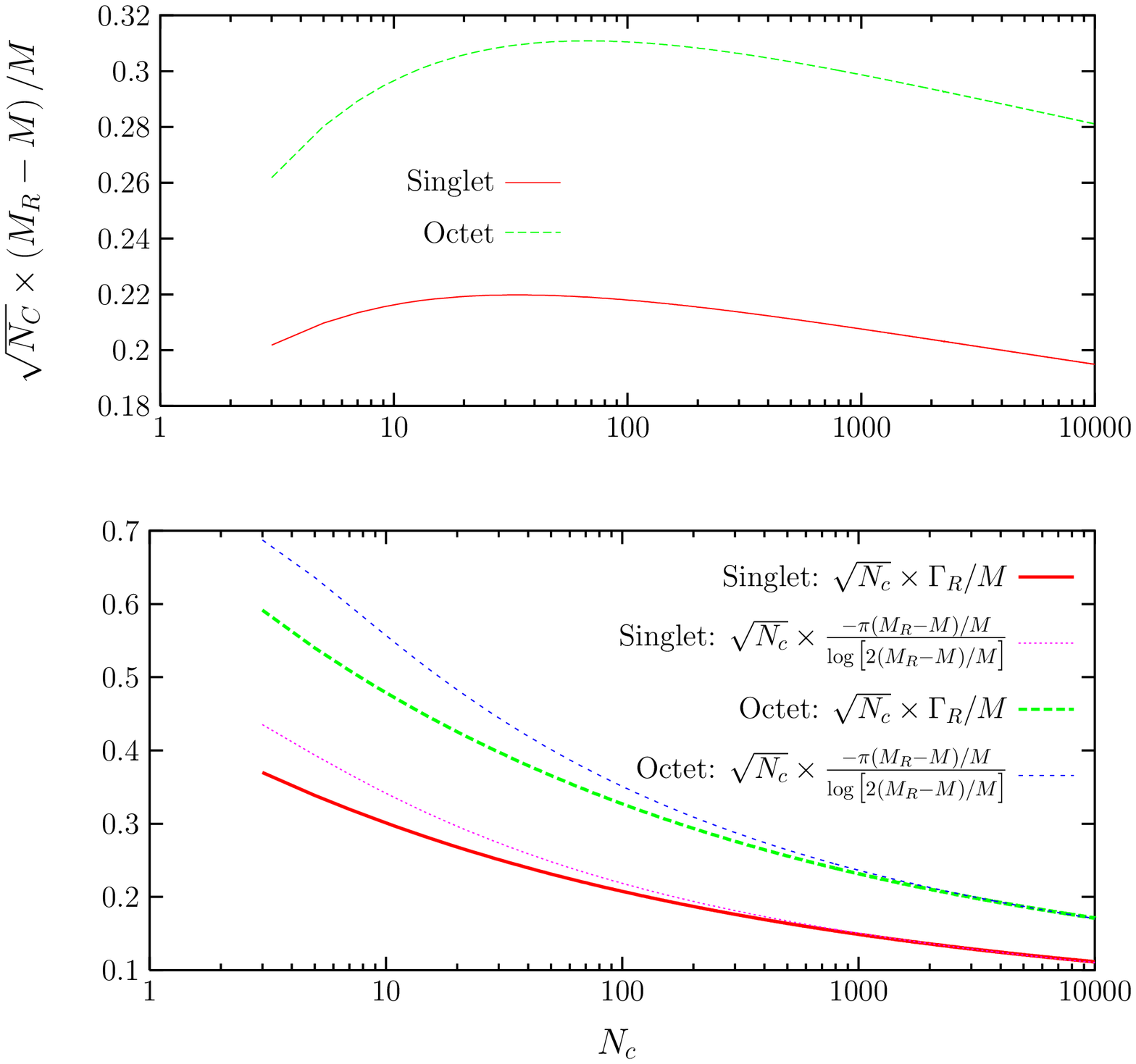}}
\vspace{-8cm}
\caption{ Singlet and octet resonance masses ($M_R$) and widths
  ($\Gamma_R$) as a function of $N_c$ from the naive large $N_c$ limit of
  the $\chi-$BS(3).}
\label{fig:fig1}
\end{figure}

Octet and singlet resonance masses and widths from
Eq.~(\ref{eq:beta}) are depicted in Fig.~\ref{fig:fig1}. 
Several comments are in order:
\begin{itemize}
\item Since $M$ increases as $N_c$, the shift $M_R-M$ and the resonance
   width, $\Gamma_R$, increase with $N_c$ 
  slower than $\sqrt{N_c}$.
\item The ratio $\Gamma_R/(M_R-M)$ approaches to zero as $N_c$ increases, both
  for singlet and octet resonances.

\item The approximate formula
\begin{equation}
\frac{\Gamma_R}{M} = - \frac{\pi \delta}{\log(2\delta)}, \quad
     {\rm with} \quad
\delta \equiv \frac{M_R-M}{M} \label{eq:apprx}
\end{equation}
works notably well in the large $N_c$ limit. Indeed, in the limit
$N_c\to\infty$ one easily finds
\begin{eqnarray}
\delta^2 \log\delta &=& \frac{24\pi^2f_0^2}{N_c\lambda_iM_0^2}
\label{eq:relsu3a} \\
\frac{\Gamma_R}{M}&=& -\lambda_i \frac{N_c\delta^3M_0^2}{24\pi
  f_0^2}, \quad i= 8,1 \label{eq:relsu3b}
\end{eqnarray}
which suggest a large $N_c$  behavior of the type
\begin{eqnarray}
\delta &\sim& \frac{1}{\sqrt{N_c\log N_c}} \label{eq:loga}\\
\frac{\Gamma_R}{M} &\sim& \frac{1}{\sqrt{N_c\log^3 N_c}} \label{eq:logb}
\end{eqnarray}
\item The presence of logarithms of $N_c$ in the mass and width of the
resonances is against standard large $N_c$ counting rules
\cite{Wi79}. In the present approach it comes out from the baryon mass
in the loop function. Such logarithms are almost certainly an artifact
of the implementation of the effective theory, and are expected to
dissappear using a more appropriate treatment along the lines of Heavy
Baryon Chiral Perturbation Theory \cite{Jenkins:1990jv}, or the
Infrared Regularization of Ellis and Tang \cite{Ellis:1997kc}, and
Becher and Leutwyler \cite{Becher:1999he}. So the BSE approach as used
in this work should reliably predict the power-like part of the $N_c$
dependence but not necessarily logarithmic corrections.

\end{itemize}
The original work of Witten~\cite{Wi79} pointed out that the excited
baryons have both natural widths and excitation energies of order
${\cal O} (N^0_c)$. More recently, some questions have been raised
about the general validity of that result and some arguments in favor
of the existence of narrow (widths of ${\cal O}(1/N_c)$) excited
baryons at large $N_c$ have been given~\cite{PY98}. Nevertheless, it
seems that a general large $N_c$ QCD analysis does not predict such
narrow states~\cite{Co04}, which has been also corroborated by other
authors~\cite{GSS05}. On the other hand, resonances are unstable
particles, and one may question the validity of a Hamiltonian
formalism\footnote{ In such scenario resonances are described as
single--quark orbital excitations about a closed-shell
core~\protect\cite{GSS02,Ope}.}, since it must be assumed that the
resonant states exist for a sufficiently long time in order to be
described as eigenstates of a Hamiltonian.  Chiral soliton models,
such as the Skyrme model, improve on that point and in those models,
resonances show up as poles in meson--baryon scattering
amplitudes~\cite{Solitons}. Recently, it has been proved that both
schemes are compatible in some sense, and give rise to a set of
multiplets of degenerate states, for which any complete spin-flavor
multiplet within one picture fills the quantum numbers of complete 
multiplets in the other picture~\cite{CL03}.

 The results of this section do not support the existence of narrow
states either, but lead to widths and excitation energies $(M_R-M)$
which do not behave as order ${\cal O} (N^0_c)$, but instead grow, in
the $N_c \to \infty$ limit, as $\sqrt{N_c}$ (modulo subleading
logarithmic corrections not under control in the present BSE
treatment). It might point out to a serious deficiency of the present
analysis.  Indeed, as we will show below, the results presented in
this section are not reliable. At least, there are two aspects which
should be revised. First, as mentioned in the introduction baryon
decuplet degrees of freedom should be included. Second, baryons carry
the quantum numbers of $N_c$ quarks (in order to form an SU($N_c$)
color singlet from color--fundamental irreps), and therefore the
baryon SU(3) irreps might depend on $N_c$, which could induce an $N_c$
dependence of the eigenvalues ($\lambda$'s). As we will see, the
extension of spin-flavor symmetry to the meson sector will also be
essential.

\section{Baryon Decuplet and Vector Meson Nonet Effects}

\label{sec:sec4}

Let us start revising the chiral Bethe Salpeter approach to SU(6)
meson-baryon scattering ($\chi-$BS(6)) developed in
Ref.\cite{GNS05}. For ground state baryons, there exists an exact
spin--flavor symmetry in the large $N_c$ limit~\cite{Ma99}. This is to
say that the light quark--light quark interaction is approximately
spin independent as well as SU(3) independent. This corresponds to
treating the six states of a light quark ($u$, $d$ or $s$ with spin
up, $\uparrow$, or down, $\downarrow$) as equivalent, and leads us to
the invariance group SU(6).  Since the pure SU(3) transformations
commute with the pure SU(2) (spin) transformations within SU(6), it
follows that a SU(6) multiplet can be decomposed into SU(3) multiplets
each of definite total spin.  With the inclusion of the spin there are
216 three quark states, and the SU(6) group representation reduction
(denoting the SU(6) multiplets by their dimensionality and a SU(3)
multiplet $\mu$ of spin $J$ by $\mu_{2J+1}$) reads 
\begin{eqnarray}
6\otimes 6 \otimes 6 &=& 56 \oplus 70 \oplus 70 \oplus 20=
\underbrace{8_2 \oplus 10_4}_{56} \oplus \underbrace{1_4\oplus
  8_2\oplus}_{20} \nonumber\\
&\oplus& 2\times \Big\{\underbrace{10_2 \oplus 8_4\oplus
  8_2\oplus 1_2 }_{70} \Big\} 
\end{eqnarray}
It is natural to assign the lowest--lying baryons to the 56--plet of SU(6),
since it can accommodate an octet of spin--$1/2$ baryons and a
decuplet of spin--$3/2$ baryons, which are exactly the SU(3)--spin
combinations of the low--lying baryon states ($(N,\Sigma,\Lambda,
\Xi)$ and ($\Delta$, $\Sigma^*$, $\Xi^*$, $\Omega$)). Furthermore, the
56--plet of SU(6) is totally symmetric, which allows the baryon to be made
of three quarks in $s$-wave. Color degrees of freedom take care of the
Fermi's statistics.

In the meson sector, assuming that the lowest lying states are
obtained from $s$-wave quark--antiquark interactions and taking into
account  the
group reduction
\begin{equation}
6\otimes 6^* = 35 \oplus 1 = \underbrace{8_1 \oplus 8_3
\oplus 1_3}_{35} \oplus \underbrace{1_1}_{1},
\end{equation}
the octet of pseudoscalar ($K, \pi,\eta, {\bar K}$) and the nonet of
vector ($K^*, \rho,\omega, {\bar K}^{*}, \phi$) mesons are commonly
placed in the 35 representation of SU(6). A ninth $0^-$ meson
($\eta^\prime$) must go in the $1$ of SU(6). The nonet of vector
mesons and the octet of Goldstone bosons are clearly not
degenerated. As mentioned in the introduction, spin-flavor symmetry in
the meson sector is not a direct consequence of large $N_c$. However,
vector mesons do exist, they will couple to baryons and presumably
will influence the properties of the resonances. Since the splitting
between the pseudo-scalar and vector mesons is of order $N_c^0$ as the
meson masses themselves, and having neglected these latter ones with
respect to the baryon masses, it is not unreasonable to assume a
spin-flavor symmetry in large $N_c$ in the meson sector, as
well. Lacking better theoretical founded models to take into account
vector mesons, studying the spin-flavor symmetric scenario seems a
reasonable first step. Moreover an underlying static chiral $U(6)
\times U(6)$ symmetry has been advocated by Caldi and
Pagels~\cite{CP76} in which vector mesons would be ``dormant''
Goldstone bosons acquiring mass thorough relativistic
corrections. This scheme solves a number of theoretical problems in
the classification of mesons and also makes predictions which are in
remarkable agreement with the experiment.

Thus for consistency, the spin--3/2 decuplet baryon and the vector
meson nonet degrees of freedom have to be added to the resonance
analysis carried out in the previous section.  As a consequence, for a given
sector ($JIS$), there  now appear  some more coupled channels than
when the involved hadrons were only the Goldstone pseudoscalar meson 
and the lowest $J^P=\frac12^+$ baryon octets. For instance, in the
$(JIS)=(1/2,0,-1)$ sector, besides the $\pi \Sigma$ , $\eta
\Lambda$, $\bar K N$ and $K \Xi$  channels, we also consider now the
$K^*\Xi$, $K^*\Xi^*$, $\rho\Sigma$, $\rho\Sigma^*$, $\omega \Lambda$,
$\bar K^* N$ and $\phi \Lambda$ ones.

We will limit ourselves to $s$-wave meson--baryon
resonances\footnote{We are aware of possible $d$-wave mixings, which
  within the framework outlined in Ref.~\protect\cite{GNS05} will be
  examined elsewhere.} and we
will make use of the SU(6) extension of the Weinberg-Tomozawa (WT)
meson-baryon chiral Lagrangian recently carried out in
Ref.~\cite{GNS05}. Chiral Symmetry (CS) at leading order (WT Lagrangian) is
much more predictive than SU(3) symmetry\footnote{From the SU(3)
decomposition of Eq.~(\ref{eq:su3}), one easily
deduces~\protect\cite{GNS05} that SU(3) symmetry describes the
Goldstone pseudoscalar meson and the lowest $J^P=\frac12^+$ baryon
octets $s$-wave scattering in terms of seven undetermined functions
(Wigner--Eckart matrix elements) of the meson--baryon Mandelstam
variable $s$.} and determines the on-shell interaction kernel,
$V(\sqrt{s})$, for ($8_1$)meson--($8_2$)baryon $s$-wave scattering in
Eq.~(\ref{eq:scat-eq}) in terms of a unique parameter ($f$), besides
the hadron masses (see Eq.~(\ref{eq:lowest})). From an SU(6) point of
view, one should work with $s$-wave meson--baryon states, constructed
out of the SU(6) 35 (mesons) and 56 (baryons) multiplets. The SU(6)
decomposition  yields
\begin{eqnarray}
\label{eq:su6-rep}
35 \otimes 56 = 56 \oplus 70 \oplus 700 \oplus 1134, 
\end{eqnarray}
and thus one has four (Wigner-Eckart irreducible matrix elements of
the SU(6) invariant Hamiltonian) free functions of the meson--baryon
Mandelstam variable $s$. It is clear that not all SU(3) invariant
interactions in the $(8_1)$meson--$(8_2)$baryon sector can be extended
to a SU(6) invariant interaction. Remarkably, the WT interaction turns 
out to be consistent with SU(6) and, moreover, the extension is
unique. In other words, there is a choice of the four couplings for
the $35 \otimes 56$ interaction that, when restricted to the $8_1
\otimes 8_2$ sector, reproduces the WT on-shell interaction kernel
$V(\sqrt{s})$ of Eq.~(\ref{eq:lowest}), and such choice is
unique~\cite{GNS05}. Indeed, the {\it potential} of
Eq.~(\ref{eq:lowest}) can be recovered, in the SU(3) limit, by taking
\begin{eqnarray}
 \langle {\cal M}^\prime  {\cal B}^\prime ; JIY | V| {\cal M} {\cal B}
 ; JIY \rangle & =&\sum_{\phi_{\rm SU(6)}} \bar\lambda_{\phi_{\rm SU(6)}}
\frac{\sqrt{s}-M}{2\,f^2} {\cal
 P}_{{\cal M}{\cal B}, {\cal M}^\prime {\cal B}^\prime }^{\phi_{\rm
 SU(6)},JIY} ,\nonumber \\
{\cal
 P}_{{\cal M}{\cal B}, {\cal M}^\prime {\cal B}^\prime }^{\phi_{\rm
 SU(6)},JIY} &=& \sum_{\mu_{\rm SU(3)},\alpha} 
\left( \begin{array}{cc|c} 35& 56&
   \phi_{\rm SU(6)} \\ \mu_M J_M & \mu_B J_B &
   \mu_{\rm SU(3)} J \alpha \end{array}\right)
\left( \begin{array}{cc|c}
 \mu_M& \mu_B& \mu_{\rm SU(3)} \\ 
  I_M Y_M & I_B Y_B & I Y\end{array}\right)
\nonumber\\
&\times&
\left( \begin{array}{cc|c} \mu^\prime_{M^\prime}& \mu_{B^\prime}^\prime&
   \mu_{\rm SU(3)} \\ I^\prime_{M^\prime}Y^\prime_{M^\prime} &
 I^\prime_{B^\prime} Y^\prime_{B^\prime} &
   IY\end{array}\right)  \left( \begin{array}{cc|c} 35& 56&
   \phi_{\rm SU(6)} \\ \mu'_{M'} J'_{M'} & \mu'_{B'} J'_{B'} &
   \mu_{\rm SU(3)} J \alpha \end{array}\right). \label{eq:vsu6}
\label{eq:su6}%
\end{eqnarray}
with
\begin{equation} 
\bar\lambda_{56}=-12, \qquad  \bar\lambda_{70}=-18, \qquad
\bar\lambda_{700}=6, \qquad \bar\lambda_{1134}=-2  \label{eq:lassu6}
\end{equation}
and $M$ now being the common octet and decuplet baryon mass. Besides,
$Y$ stands for the hypercharge (strangeness plus baryon number), and
we use the notation ${\cal M}\equiv \left [(\mu_M)_{2J_M+1}, I_M,
Y_M\right]$ for mesons and similarly for baryons (${\cal B}$). Thus,
$\mu_{M}=8,1$ and $\mu_{B}=8,10$ are the meson and baryon SU(3)
multiplets, respectively, and $J_{M,B},I_{M,B},Y_{M,B}$ are the spin,
isospin and hypercharge quantum numbers of the involved
hadrons. Finally in Eq.~(\ref{eq:su6}), SU(3) isoscalar
factors~\cite{CG-su3}, and the SU(6)--multiplet coupling
factors~\cite{CG-su6} are also used. For more details see
Ref.~\cite{GNS05}.

\subsection{Naive large $N_c$ limit of the $\chi-$BS(6)}

\label{sec:sec4a}

In this subsection we use Eq.~(\ref{eq:vsu6}) as the interaction
kernel to solve the BSE in the large $N_c$ limit. Thus, we improve on
the analysis of Sect.~\ref{sec:sec3} by including baryon decuplet and
vector meson nonet effects. We assume that the ground state baryons
fall in the 56--plet of SU(6) (only possible if $N_c$ is odd) for
large $N_c$, however we still ignore the fact that the spin-flavor
irreps might depend on $N_c$.  

Since $V$ is a SU(6) scalar operator, one readily realizes that the
resonance equation reads now,
\begin{equation}
\beta(s)\Big|_{s=s_R\equiv M^2_R-iM_R\Gamma_R} =  \bar 
\lambda_{\phi_{SU(6)}}, \quad
\phi_{SU(6)}=56,70,700,1134
\label{eq:beta_SU(6)} 
\end{equation} 
with $M_R>M$ and $\Gamma_R>0$ and thus, the approximated relation of
Eq.~(\ref{eq:apprx}) is accomplished, as well. This equation has
solutions only for negative eigenvalues, $\bar \lambda_{70}$, $\bar
\lambda_{56}$ and $\bar \lambda_{1134}$.  Note that the 70 of SU(6)
leads to the most attractive $s$-wave meson--baryon interaction. This is
also the scenario commonly adopted in most large $N_c$ works,
where the first negative parity baryon excited states are considered
as members of the 70 multiplet (see f.i. Ref~\cite{GSS02}).  Beyond
the large $N_c$ LO, there appear terms in the meson--baryon
Hamiltonian which explicitly break down the spin--flavor symmetry and
thus, one expects SU(6) configuration mixings. Hence the $N(1535)$,
$N(1650)$, $\Lambda(1405)$, $\Lambda(1670)$, $\Sigma(1620)$,
$\Xi(1620)$, $\Xi(1690)$, $\Lambda(1390)$ $s$-wave $\frac 12^-$
resonances will be constructed out of the SU(6) 70, 56 and 1134
resonant states.  From the spin--flavor content of the SU(6)
representations, we expect the SU(3) singlet resonance\footnote{The
resonance $\Lambda(1390)$ will have a large SU(3) singlet component.}
to be a linear combination of the 70 and 1134 states, while for the
SU(3) octet ones, the SU(6) 56 resonant states will have to be
considered as well\footnote{The 56-plet should be included, since
there is only one $8_2$ multiplet in the 70 of SU(6).}.

Thus, the properties of these resonances ($N(1535)$, $N(1650)$,
$\Lambda(1405)$,...) are modified by their coupling to the baryon
decuplet and vector meson nonet states. Assuming that these states
belong to the 70~\cite{GSS02}\footnote{The available analysis of the
negative parity 70--plet baryon masses within the $1/N_c$ expansion
suffer from a serious deficiency. Those studies do not consider the
$\Lambda(1390)$ state, which existence has been firmly established
from a theoretical point of view~\protect\cite{Ga03,nrg,Ji03}, and
also there are some indications supporting its existence in the $K^- p
\to \pi^0 \pi^0 \Sigma^0$ reaction data, as it has been recently
pointed out in Ref.~\cite{ORM05}. Traditionally, large $N_c$ studies
construct the isospin singlet states of the $8_2$ and $1_2$
SU(3)$_{2J+1}$ representations, entering in the 70 SU(6) multiplet, as
linear combinations of the $\Lambda(1405)$ and $\Lambda(1670)$
resonances. It is clear, that the $\Lambda(1390)$ state should be
considered, and presumably it will have a large $1_2$ component.}, 56
and 1134 multiplets, we find that the relations of
Eqs.~(\ref{eq:relsu3a})--(\ref{eq:relsu3b}) are correct, just by
replacing $\lambda_{1,8}$ by $\bar \lambda_{70,56,1134}$.

On the other hand, the relations of
Eqs.~(\ref{eq:loga})--(\ref{eq:logb}) still hold, with no
modifications. As a conclusion, considering baryon decuplet and vector
meson nonet effects would lead to some quantitative changes on
resonance masses and widths at relatively low values of $N_c$, and
would affect to the rate how the $N_c \to \infty$ relations of
Eqs.~(\ref{eq:loga})--(\ref{eq:logb}) are reached. Therefore, widths
and excitation energies $(M_R-M)$ would not behave as order ${\cal O}
(N^0_c)$, but they would still grow as $\sqrt{N_c}$. Thus, the
inclusion of baryon decuplet and vector meson nonet degrees of
freedom, treated as in this subsection, does not modify this behavior,
possibly incorrect.

In the analysis presented up to here, it has been ignored the fact
that, since baryons for arbitrary $N_c$ contain $N_c$ valence quarks,
the corresponding baryon SU(6) representations also grow in size with
$N_c$~\cite{DJM94}. As we will see in the next subsection, this will
provide an explicit $N_c$ dependence for the eigenvalues
$\bar\lambda_{\phi_{SU(6)}}$. This further $N_c$ dependence will allow
us first to show that, in some SU(6) irreducible spaces, the SU(6)
extension of the WT $s$-wave meson-baryon interaction, sketched in this
subsection, scales as ${\cal O}(N_c^0)$, instead of the well known
${\cal O}(N_c^{-1})$ behavior for its SU(3) counterpart, and second to
recover the Witten's scaling rules for both widths and excitation
energies of the resonant states.

\subsection{ Extension of the $\chi-$BS(6) Model for  Arbitrary $N_c$}
\label{sec:nc}

\subsubsection{ SU(6) representations and WT Lagrangian for  Arbitrary
  $N_c$ } 
\label{sec:wtnc}

Mesons at arbitrary $N_c$ still carry the quantum numbers of a single
$q{\bar q}$, and hence their SU(2$N_F$) spin-flavor irreps are
unchanged when $N_c$ is changed. Thus, the octet of pseudoscalar ($K,
\pi,\eta, {\bar K}$) and the nonet of vector ($K^*, \rho,\omega, {\bar
K}^{*}, \phi$) mesons are placed in the 35 representation of
SU(6). Baryons, on the other hand, carry the quantum numbers of $N_c$
quarks (in order to form an SU($N_c$) color singlet from
color--fundamental irreps), and therefore the baryon SU(2$N_F$)
spin-flavor irreps grow in size with $N_c$. We wish to identify these
large $N_c$ representations with their $N_c=3$ counterparts. As it is
done in Ref.~\cite{CL04}, to keep our notation simple and aid in the
extrapolation to three colors case, we use quotes to denote the
generalized SU(2$N_F$) representations familiar from three colors. The
ground-state spin-flavor multiplet is taken to be completely symmetric
$N_c-$tableau representation, which is the analog to the SU(6) 56 for
three flavors,
\begin{equation}
\underbrace{\begin{Young} & & & & \cr \end{Young} \ldots \begin{Young}
    & & & & \cr \end{Young} }_{N_c\, {\rm boxes}}
\end{equation}
Notationally, we denote such arbitrary-$N_c$ generalization as ``56'' 
and its dimension is $\left (\begin{array}{c}N_c+5\cr
5\end{array}\right)$. The SU(6) decomposition of Eq.~(\ref{eq:su6-rep})
now, for arbitrary $N_c$,  reads
\begin{eqnarray}
35 \otimes \makebox{\rm ``56''}& \equiv & 
 \begin{Young} &  \cr\cr\cr\cr\cr \end{Young}~\otimes~ \overbrace{\begin{Young} & & & & \cr
    \end{Young}}^{N_c} \nonumber\\
&=&
\overbrace{\begin{Young} & & & & \cr
    \end{Young}}^{N_c}~  \oplus~
    \overbrace{\begin{Young} & & & & \cr \cr\end{Young}}^{N_c-1} ~\oplus~\overbrace{\begin{Young} & & & & \cr\cr\cr\cr\cr
    \end{Young}}^{N_c+2} ~\oplus~ \overbrace{\begin{Young} & & & & \cr&\cr\cr\cr\cr \end{Young}}^{N_c+1} \nonumber \\&&\nonumber \\
&=& \makebox{\rm ``56''} \oplus \makebox{\rm ``70''} \oplus
    \makebox{\rm ``700''} \oplus \makebox{\rm ``1134''}, \label{eq:su6_nc} 
\end{eqnarray}
where the dimensions of the ``70'', ``700'' and ``1134'' irreps are
$\frac{5(N_c-1)}{(N_c+5)}\left (\begin{array}{c}N_c+5\cr 5\end{array}\right)$,
$\frac{5(N_c+7)}{(N_c+1)}\left (\begin{array}{c}N_c+5\cr 5\end{array}\right)$
and $\frac{24N_c(N_c+6)}{(N_c+5)(N_c+1)}\left (\begin{array}{c}N_c+5\cr
5\end{array}\right)$, respectively. Thus,
we find a first remarkable result: assuming SU(6) spin-flavor
symmetry, the $s$-wave 35--meson ``56''--baryon scattering for an arbitrary
value of $N_c$, can still be described in terms of four (Wigner-Eckart
irreducible matrix elements of the SU(6) invariant Hamiltonian)
undetermined functions of the meson--baryon Mandelstam variable
$s$. This is also the case for any number of flavors $N_F\ge 2$. 

Next step is to make use of the underlying CS to
further constrain these four undetermined functions. For SU(3) flavor
symmetry and $N_c=3$, the latter functions, at LO in the
chiral expansion, are determined by the WT Lagrangian. It is not just
SU(3) symmetric but also chiral
($\text{SU}_L(3)\otimes\text{SU}_R(3)$) invariant. Symbolically,
\begin{equation}
{\cal L_{\rm WT}}= {\rm Tr} ( [M^\dagger, M]B^\dagger B)
\end{equation}
This structure, dictated by CS, is more suitably analyzed in the
$t$-channel. The meson, $M$, and baryon, $B$, fields fall in the representation
SU(3) $8$ which is also the adjoint representation. The commutator
$[M^\dagger,M]$ indicates a $t$-channel coupling to the $8_a$
(antisymmetric) representation, thus
\begin{equation}
{\cal L_{\rm WT, SU(3)}}= \left((M^\dagger\otimes M)_{8_a}\otimes
(B^\dagger\otimes B)_{8}\right)_{1}
\end{equation}
The unique SU(6) extension is then
\begin{equation}
{\cal L_{\rm WT, SU(6)}}= \left((M^\dagger\otimes M)_{35_a}\otimes
 (B^\dagger\otimes B)_{35}\right)_{1}
\label{eq:1}
\end{equation}
since the 35 is the adjoint representation of SU(6). The $t$-channel
decompositions $35 \otimes 35 = 1 \oplus 35_s\oplus 35_a\oplus 189
\oplus 280 \oplus 280^* \oplus 405$ and $56 \otimes 56^* = 1 \oplus 35
\oplus 405 \oplus 2695$ indicate that the coupling in Eq.~(\ref{eq:1})
exists and is indeed unique~\cite{GNS05}, all coupling constants being
reduced to a single independent one, namely, that of the WT Lagrangian
(pion weak decay constant, besides the hadron masses). To extend this
result to arbitrary $N_c$, we should first consider
\begin{eqnarray}
\makebox{\rm ``56''} \otimes  ``56^*\makebox{\rm ''}& \equiv & 
 \overbrace{\begin{Young} & & & & \cr
    \end{Young}}^{N_c} ~\otimes~
 \overbrace{\begin{Young} & & & & \cr & & & &\cr & & & & \cr& & & &
     \cr& & & & \cr \end{Young}}^{N_c}  \nonumber\\ &&\nonumber\\
&=&
\overbrace{\begin{Young}  &   &   &  &
      &  &  & \cr&&&\cr&&&\cr&&&\cr &&&\cr\end{Young}}^{2N_c} ~\oplus~~~\overbrace{\begin{Young}  &   &   &  &
      &  &  & \cr&&&\cr&&&\cr&&&\cr &&&\cr\end{Young}}^{2N_c-2}~\oplus~~~\overbrace{\begin{Young}  &   &   &  &
      &  &  & \cr&&&\cr&&&\cr&&&\cr &&&\cr\end{Young}}^{2N_c-4}~\oplus \ldots \oplus~~ \begin{Young} &  \cr\cr\cr\cr\cr
     \end{Young}~\oplus~ 1 \label{eq:young}\\
&{\phantom =}&
\phantom{a}^{\leftarrow N_c \rightarrow} ~~~~~~~~~~~~~~~~~~~~\,
\phantom{a}^{\leftarrow N_c-1
  \rightarrow}~~~~~~~~~~~~~~~~~~~ \phantom{a}^{\leftarrow N_c-2
  \rightarrow} \nonumber 
\end{eqnarray}
where the dimension of the tableau with $2n$ boxes in the first row is
$(2n+5)\left (\begin{array}{c}n+5\cr 5\end{array}\right)\left
  (\begin{array}{c}n+4\cr 4\end{array}\right)/(n+5)$, accomplishing
\begin{equation}
\sum_{n=0}^{N_c} \frac{2n+5}{n+5}\left (\begin{array}{c}n+5\cr
  5\end{array}\right)\left (\begin{array}{c}n+4\cr 4\end{array}\right)
  = \left (\begin{array}{c}N_c+5\cr
5\end{array}\right)^2
\end{equation}
to verify the equality between the dimensions of both sides of
Eq.~(\ref{eq:young}). We see that the SU(6) 35 (adjoint
representation) appears in the decomposition into irreps of 
``56''$\otimes ``56^*\makebox{\rm ''}$
(Eq.~(\ref{eq:young})), and thus we find that the SU(6) extension of
the WT Lagrangian (Eq.~(\ref{eq:1})) can still be done for arbitrary $N_c$.

Let us denote the contravariant and covariant spin-flavor quark  and
antiquark components

\begin{equation}
q^i = \left (\begin{array}{c}u\uparrow \cr d\uparrow \cr s\uparrow \cr
u\downarrow \cr d\downarrow \cr s\downarrow \end{array}\right ),
\qquad {\bar q}_i = \left ({\bar u}\downarrow,-{\bar
  d}\downarrow,-{\bar s}\downarrow, -{\bar u}\uparrow, {\bar
  d}\uparrow, {\bar s}\uparrow    \right)
\end{equation}
where $q^i$ (${\bar q}_i$) annihilates\footnote{We use a
convention such that $\left(\begin{array}{c}{\bar d}\cr {\bar
u}\end{array}\right)$ is a standard basis of SU(2), that is ${\bar d} =
|1/2,1/2\rangle $ and ${\bar u} = |1/2,-1/2\rangle $. Thus, ${\bar u},
{\bar d} , {\bar s}$ is a standard basis of the $3^*$ representation of
SU(3) with de Swart's convention~\cite{CG-su3}.} a quark (antiquark)
with the spin-flavor $i$. For instance ${\bar u}\downarrow$
annihilates an antiquark with flavor ${\bar u}$ and $S_z=-1/2$.
Mesons fall in the adjoint representation and we represent 
the annihilation operators of mesons in the 35 of SU(6) by means of a
traceless  tensor $M^i_j$, which under SU(6) transformations behaves
like 
\begin{equation}
 q^i{\bar q}_j - \frac{1}{2N_F} q^m{\bar q}_m \delta^i_j,\quad
i,j=1,\ldots 2N_F \label{eq:defm}
\end{equation}
with $N_F$ the number of flavors, three in this work. We represent the
annihilation operators of baryons in the ``56'' of SU(6), for arbitrary
$N_c$, by means of a completely symmetric tensor $B^{i_1i_2\ldots
i_{N_c}}$, which under SU(6) transformations behaves like
$q^{i_1}q^{i_2}\ldots q^{i_{N_c}}$. We treat $q^i$ as boson fields,
since the color wave function, not explicitly shown, is fully
antisymmetric. The corresponding Wick's contractions of these fields read
\begin{eqnarray}
\underwick{1}{<1M^k_l >1M^{\dagger i}_j}&=&
\delta^k_j\delta^i_l-\frac{1}{2N_F} \delta^i_j\delta^k_l \nonumber \\
\underwick{1}{<1B^{j_1j_2\ldots j_{N_c}}>1B^\dagger_{i_1i_2\ldots
i_{N_c}}}&=& \sum_{P\in S_{N_c}}
\delta_{i_1}^{P(j_1)}\delta_{i_2}^{P(j_2)} \ldots
\delta_{i_{N_c}}^{P(j_{N_c})} \label{eq:wick}
\end{eqnarray}
where $S_{N_c}$ is the group of permutations of $N_c$ objects and we
use a notation such that the $N_c-$tuple $P(i_1)P(i_2)\ldots
P(i_{N_c})$ is equal to $P(i_1i_2\ldots i_{N_c})$.

From the discussion above, we find that Eq.~(\ref{eq:1}) is still the
unique SU(6) extension of the WT $s$-wave meson-baryon
interaction for arbitrary $N_c$. Thus, we find that the group
structure, ${\cal G_{\rm WT,SU(6)}^{\rm SU(N_c)}}$, of the SU(6)
extension of the WT, up to constant factors, takes the
form\footnote{Here and in most of the Subsect.~\ref{sec:nc}, though we give
  explicitly expressions for the $N_F=3$ case, the formulae are
  easily extended for an arbitrary  number of flavors.}
\begin{equation}
{\cal G_{\rm WT,SU(6)}^{\rm SU(N_c)}}=   
\frac{2}{(N_c-1)!}:\left ( M^i_j M^{\dagger j}_k- M^{\dagger
i}_j M^j_k\right )B^\dagger_{ii_2\ldots
i_{N_c}}B^{ki_2\ldots i_{N_c}}:  \label{eq:gwt}
\end{equation}
where $:...:$ denotes the normal product and the factor $2/(N_c-1)!$
has been introduced for convenience. To obtain the full form of the
Hamiltonian, one should specify some constant factors
\begin{equation}
 {\cal H_{\rm
WT,SU(6)}^{\rm SU(N_c)}} \propto {\cal G_{\rm WT,SU(6)}^{\rm SU(N_c)}},
\end{equation}
which depend on kinematics and possibly also on the number of
colors. These factors will be discussed in Subsect.~\ref{sec:hwt}.

\subsubsection{ Explicit form of ${\cal H_{\rm WT,SU(6)}^{\rm SU(N_c)}}$}

\label{sec:hwt}

First and for $N_c=3$, we will write a $s$-wave meson-baryon Lagrangian
invariant under SU(3)$\times$ SU(2) transformations and involving only
the Goldstone boson and the nucleon octets. Starting from the lowest
order in the chiral expansion~\cite{Pich95}\footnote{We have
omitted the pieces proportional to the couplings ${\cal D}$ and ${\cal
F}$ (${\cal F}+{\cal D} = g_{A} = 1.25$) because they do not
lead to the WT interaction Lagrangian.}
\begin{eqnarray}
{\cal L}_1 = {\rm Tr} \left\{ \bar{\Psi}_B \left( {\rm i}
\slashchar{\nabla} - M \right) \Psi_B \right\} 
\label{eq:LB1}
\end{eqnarray}
where $M$ is the common mass of the baryon octet due to SCSB for massless
quarks and ``Tr'' stands for the trace in SU(3). In addition,
\begin{eqnarray}
\nabla^{\mu} \Psi_B &=& \partial^{\mu} \Psi_B + [A^\mu_3,\Psi_B] \nonumber \\
A^\mu_3 &=& \frac{1}{2} \left ( u_3^\dagger
\partial^{\mu} u_3 + u_3 \partial^{\mu} u_3^\dagger \right ) =
\frac{1}{4f^2}  [\Phi_3, \partial^\mu\Phi_3]+ {\cal O}((\Phi_3)^4) \nonumber \\
U_3 = u_3^2 &=& e^{ {\rm i} \sqrt{2} \Phi_3 / f } 
\end{eqnarray} 
The SU(3) matrices for the meson and the baryon octets are
written in terms of the meson and baryon Dirac fields respectively
and are given by\footnote{For the purpose of our work we do not
consider any mixing between octet and singlet SU(3) representations}
\begin{eqnarray}
	\Phi_3 = \left( \matrix{ \frac{1}{\sqrt{2}} \pi^0 +
	\frac{1}{\sqrt{6}} \eta & \pi^+ & K^+  \cr  \pi^- & -
	\frac{1}{\sqrt{2}} \pi^0 + \frac{1}{\sqrt{6}} \eta & K^0  \cr 
	K^- & \bar{K}^0 & - \frac{2}{\sqrt{6}} \eta }
	\right) \, ,
\end{eqnarray}
and 
\begin{eqnarray}
	\Psi_B =
	\left( \matrix{ 
	\frac{1}{\sqrt{2}} \Sigma^0 + \frac{1}{\sqrt{6}} \Lambda &
		\Sigma^+ & p \cr 
		\Sigma^- & - \frac{1}{\sqrt{2}} \Sigma^0 
		+ \frac{1}{\sqrt{6}} \Lambda & n \cr 
		\Xi^- & \Xi^0 & - \frac{2}{\sqrt{6}} \Lambda } 
	\right) \, .
\end{eqnarray}
respectively. Performing a non-relativistic reduction\footnote{In
  order to find an SU(6) invariant Lagrangian, it is natural to
  perform an non-relativistic reduction, since the no-go
  Coleman-Mandula theorem~\cite{CM67} forbids an exact hybrid symmetry
  mixing a compact internal flavor symmetry with the non-compact Poincare
  symmetry of spin angular momentum. Furthermore, in the large $N_c$
  limit,  a non-relativistic treatment of
  baryons is totally justified.} of Eq.~(\ref{eq:LB1}), we
  find
\begin{equation}
{\cal L}_{1\, \rm norel} =  {\rm Tr} \left\{
B^\dagger_3 \left( {\rm i} \nabla_0 - M + \frac{1}{2M}
(\vec{\sigma}\cdot\vec{\nabla})^2 \right) B_3 \right\} 
\end{equation}
where now the $B_3-$fields (large components of the $\Psi_B$ ones) do
not contain antiparticle degrees of freedom, that is they are
bispinors which account for the spin degrees of freedom of the
non-relativistic baryons. The above Lagrangian is not invariant under
SU(3)$\times$ SU(2) transformations, yet. This is because of a
spin-orbit type interaction generated by the Pauli matrices. Such a
term does not contribute to $s$-wave and neglecting it, we get
\begin{equation}
{\cal L}_{{\rm SU(3)}\times {\rm SU(2)} } = {\rm Tr} \left\{
B^\dagger_3 \left( {\rm i} \nabla_0 - M + \frac{1}{2M}
\vec{\nabla}^2 \right) B_3 \right\} \label{eq:su3su2}
\end{equation}
which is now  SU(3)$\times$SU(2) invariant. Neglecting non $s$-wave
contributions and including explicit baryon mass breaking terms, the
interaction part of the above Lagrangian leads to the chiral LO amplitude  of
Eq.~(\ref{eq:lowest}). 

The extension of the Lagrangian of Eq.~(\ref{eq:su3su2}) to describe
also baryon decuplet and vector meson nonet degrees of freedom is
now straightforward and it reads
\begin{equation}
{\cal L}_{{\rm SU(6)}} = {\rm Tr} \left\{
B^\dagger_6 \left( {\rm i} \nabla_0 - M + \frac{1}{2M}
\vec{\nabla}^2 \right) B_6 \right\} 
\label{eq:lawt} 
\end{equation}
where ``Tr'' stands now for the trace in SU(6). In addition,
\begin{eqnarray}
\nabla^{\mu} B_6 &=& \partial^{\mu} B_6 + A^\mu_6*B_6 \nonumber \\
A^\mu_6 &=& \frac{1}{2} \left ( u_6^\dagger
\partial^{\mu} u_6 + u_6 \partial^{\mu} u_6^\dagger \right )= 
\frac{1}{4f^2_6} [\Phi_6, \partial^\mu\Phi_6]+ {\cal
O}((\Phi_6)^4) \nonumber \\ U_6 = u_6^2 &=& e^{ {\rm i} \sqrt{2}
\Phi_6 / f_6 }
\end{eqnarray} 
$M$ is now the common mass of the 56 baryon representation and $f_6=
f/\sqrt{2}$, as shown in Appendix~\ref{sec:restr}. Besides, $B_6$ and
$\Phi_6$ are the baryon and meson fields which now belong to the 56
and 35 irreps of SU(6), respectively and the meaning of $A^\mu_6*B_6$
will be specified later (see Eq.~(\ref{eq:cov1}))\footnote{For the
SU(3) case $A^\mu_3*B_3$ reduces to the usual commutator. For SU(6),
it will not be a commutator since while $A^\mu_6$ are dimension six
traceless matrices, the $B_6$ baryon field is a fully symmetric tensor
with $N_c$ indices.}.  Obviously, we need to check that the
restriction of the above Lagrangian to the $8_1 \otimes 8_2$ sector
reproduces that given in Eq.~(\ref{eq:su3su2}). This is explicitly
shown in Appendix~\ref{sec:restr}, for three flavors, though the
extension to $N_F$ flavors is straightforward.

In the above equations, $\Phi_6$ is a dimension six matrix made of
full meson fields, which depend on the space-time coordinates.  The
annihilation part of the meson matrix $[\Phi_6]^i_j$ is determined by
the operators $M^i_j$ (see Eq.~(\ref{eq:defm})). On the other hand,
for SU(6) and arbitrary $N_c$, we will work with baryon fields ${\cal
B}^{i_1i_2\ldots i_{N_c}}$ such that their Fock space structure is
determined by the operators $B^{i_1i_2\ldots i_{N_c}}$ introduced in
Subsect.~\ref{sec:wtnc}. Thus, we have\footnote{With this convention
${\cal B}^{123\ldots N_c}= {\cal B}^{213\ldots N_c} = \ldots $ or
${\cal B}^{111\ldots 1} / \sqrt{N_c!} $ are baryon fields with the
usual normalization. This is because
\begin{eqnarray}
\underwick{1}{<1B^{123\ldots N_c}>1B^\dagger_{123\ldots N_c}} &=&
1,\qquad {\rm while} \nonumber \\
\underwick{1}{<1B^{111\ldots 1}>1B^\dagger_{111\ldots 1}} &=& N_c!
\end{eqnarray}
Thus, for instance for $N_c=3$,  ${\cal B}^{111}/\sqrt{3!} =
\Delta^{++}(S_z=+3/2)$.}
\begin{equation}
\frac{1}{N_c!} {\cal B}^\dagger_{i_1i_2\ldots i_{N_c}} {\cal
  B}^{i_1i_2\ldots i_{N_c}} = \sum_{\lambda \in \makebox{\rm ``56''}}
  b^\dagger_\lambda b_\lambda, \qquad i_1,i_2,\ldots, i_{N_c}\in \{1,\ldots
  ,2N_F\} \label{eq:su6norm}
\end{equation}
with $b_\lambda$ a ``56'' baryon field. In terms of these baryon fields the extension of the Lagrangian of
Eq.~(\ref{eq:lawt}) for an arbitrary number of colors
reads
\begin{equation}
{\cal L}_{{\rm SU(6)}}^{\rm SU(N_c)} = 
\frac{1}{N_c!} {\cal B}^\dagger_{i_1i_2\ldots i_{N_c}} 
 \left( {\rm i} \nabla_0 - M + \frac{1}{2M}
\vec{\nabla}^2 \right) {\cal B}^{i_1i_2\ldots i_{N_c}}, \qquad
i_1,i_2,\ldots, i_{N_c} \in \{1,\ldots ,2N_F\} \label{eq:wtnc2}
\end{equation}
where the covariant derivative acts on the baryon fields ${\cal B}$ as
usual
\begin{equation}
\left(\nabla^\mu {\cal B} \right)^{i_1i_2\ldots i_{N_c}} =
\left(\partial^\mu {\cal B} + A^\mu_6*{\cal B}\right)^{i_1i_2\ldots
  i_{N_c}} = 
\partial^\mu {\cal B}^{i_1i_2\ldots i_{N_c}} + [A_6^\mu]^{i_1}_k{\cal
  B}^{ki_2\ldots i_{N_c}} +\ldots + [A_6^\mu]^{i_{N_c}}_k{\cal
  B}^{i_1i_2\ldots i_{N_c-1}k} \label{eq:cov1} \\
\end{equation}
and therefore, we find thanks to the symmetry of 
the baryonic tensor ${\cal B}^{i_1i_2\ldots i_{N_c}}$
\begin{equation}
{\cal B}^\dagger_{i_1i_2\ldots i_{N_c}} \left(\nabla^\mu {\cal B}
\right)^{i_1i_2\ldots i_{N_c}} = {\cal B}^\dagger_{i_1i_2\ldots
  i_{N_c}} \left( \partial^\mu {\cal B}^{i_1i_2\ldots
i_{N_c}} + N_c [A_6^\mu]^{i_1}_k{\cal B}^{ki_2\ldots i_{N_c}}\right)
\label{eq:cov}
\end{equation}

From Eqs.~(\ref{eq:wtnc2}) and~(\ref{eq:cov}) we get\footnote{We have
replaced $\vec{\nabla}^2$ by $\vec{\partial}^{\,2}$ since the
difference between both operators does not contribute to $s$-wave.}
\begin{eqnarray}
{\cal L}_{{\rm SU(6)}}^{\rm SU(N_c)} & = & {\cal L}_{{\rm kin,\,
SU(6)}}^{\rm SU(N_c)}+ {\cal L}_{{\rm WT,\,SU(6)}}^{\rm SU(N_c)} \\ \nonumber\\
{\cal L}_{{\rm kin,\,
SU(6)}}^{\rm SU(N_c)} &=& \sum_{\lambda \in \makebox{\rm ``56''}}
  b^\dagger_\lambda \left( {\rm i} \partial_0 - M + \frac{1}{2M}
\vec{\partial}^{\,2} \right) b_\lambda \\ \nonumber\\
{\cal L}_{{\rm WT,\,SU(6)}}^{\rm SU(N_c)}  
&=& \frac{{\rm i} N_c}{2f^2} [\Phi_6, \partial_0 \Phi_6]^j_k 
\frac{1}{N_c!} {\cal
  B}^\dagger_{ji_2\ldots i_{N_c}} {\cal
  B}^{ki_2\ldots i_{N_c}}
\end{eqnarray}
which corresponds to the decomposition of ${\cal L}_{{\rm
SU(6)}}^{\rm SU(N_c)}$ into a baryon kinetic and an $s$-wave
meson-baryon interaction (WT) terms. The WT Hamiltonian\footnote{This
is an abuse of notation, we really mean the on-shell scattering
amplitude, at LO in the chiral expansion, which  is used as the
kernel, $V$, for the on-shell BSE.}  $\left ({\cal H_{\rm
WT,SU(6)}^{\rm SU(N_c)}}= -{\cal L_{\rm WT,SU(6)}^{\rm
SU(N_c)}}\right)$, acting on meson-baryon Fock states $|r\rangle$,
takes the form
\begin{equation}
{\cal H_{\rm WT,SU(6)}^{\rm SU(N_c)}} |r\rangle =
    \frac{\sqrt{s}-M}{2f^2}\times  {\cal G_{\rm
    WT,SU(6)}^{\rm SU(N_c)}}|r\rangle 
\end{equation} 
with ${\cal G_{\rm WT,SU(6)}^{\rm SU(N_c)}}$ defined in
Eq.~(\ref{eq:gwt}). From the results of Appendix~\ref{sec:gwt}, we
conclude that ${\cal H_{\rm WT,SU(6)}^{\rm SU(N_c)}}$ is diagonal in
the spaces associated to the ``56'', ``70'', ``700'' and ``1134''
representation of SU(6) and with eigenvalues:
\begin{equation}
\bar\lambda_{\makebox{\rm \footnotesize ``56''}} = -4N_F,
\qquad \bar\lambda_{\makebox{\rm \footnotesize ``70''}} =
-2(N_c+2N_F), \qquad \bar\lambda_{\makebox{\rm \footnotesize ``700''}}
= 2N_c, \qquad \bar\lambda_{\makebox{\rm
\footnotesize ``1134''}} =- 2
\end{equation} 
Note that for the case $N_F=3$ and $N_c=3$, we nicely recover
$\bar\lambda_{56}=-12$, $\bar\lambda_{70}=-18$, $\bar\lambda_{700}=6$
and $\bar\lambda_{1134}=-2$ (Eq.~(\ref{eq:lassu6})). Remarkably, we
see that in the ``70'' and ``700'' irreducible spaces, the SU(6)
extension of the WT $s$-wave meson-baryon interaction
scales as ${\cal O}(N_c^0)$, instead of the well known ${\cal
O}(N_c^{-1})$ behavior for its SU(3) counterpart. 


\subsubsection{Large $N_c$ SU(6) versus SU(3) WT interaction}

It would seem that the additional $N_c$ factor obtained here, as
compared to the standard SU(3) calculation, comes only from a proper
treatment of the baryon, namely, to use the correct $N_c$-dependent
``56'' representation instead of the 56. This is only partially true:
another crucial ingredient has been the introduction of vector mesons
in the scheme. Note that if one only considers pseudoscalar mesons, the
interaction being $s$-wave, the various baryonic spin sectors will
never mix and if one starts with nucleons, the ``decuplet'' states
will not be seen. In order to further analyzed this point, let us
write the meson field in the form
\begin{equation}
\Phi_6 = \frac{1}{2}\pi_\alpha \lambda_\alpha
+\frac{1}{2}\rho_{\alpha i}  \lambda_\alpha \sigma_i
+\frac{1}{\sqrt{2N_F}}\omega_i \sigma_i
:=
\sqrt{2}\phi_A t_A
\,.
\end{equation}
Here $\sigma_i$ and $\lambda_\alpha$ are Pauli and (su($N_F$) algebra)
Gell-Mann matrices with $i=1,2,3$ and $\alpha=1,\ldots,N_F^2-1$, and
$\pi$, $\rho$, $\omega$ are the hermitian meson fields corresponding
to the pseudoscalar octect and the vector nonet. The matrices $t_A$
are the SU($2N_F$) group  generators in the fundamental representation,
namely, $(\lambda_\alpha\otimes 1)/\sqrt{8}$,
$(1\otimes\sigma_i)/\sqrt{4N_F}$, and
$(\lambda_\alpha\otimes\sigma_i)/\sqrt{8}$, and $\phi_A$ the
associated meson fields with $A=1,\ldots,(2N_F)^2-1$. For the matrix
element $M_A B\to M_{A^\prime} B^\prime$ one then finds
\begin{eqnarray}
\frac{1}{4}\langle M_{A^\prime} B^\prime|
{\cal G_{\rm WT,SU(6)}^{\rm SU(N_c)}} |M_A B \rangle
&=&
\frac{1}{(N_c-1)!}([t_A,t_{A^\prime}])^j_{~k}\langle B^\prime| 
B^\dagger_{ji_2\ldots i_{N_c}} 
  B^{ki_2\ldots i_{N_c}}
|B\rangle
\nonumber \\
&=&
([t_A,t_{A^\prime}])^j_{~k}\langle B^\prime|q^\dagger_j q^k
|B\rangle
\nonumber \\
&=&
\langle B^\prime| [G_A,G_{A^\prime}]
|B\rangle
\label{eq:59}
\end{eqnarray}
where
\begin{equation}
G_A= (t_A)^j_{~k} \, q^\dagger_j q^k
\end{equation}
are the SU($2N_F$) generators on the baryon sector,
\begin{equation}
G_A= T_\alpha,\,S_i\,, G_{\alpha i} \,.
\end{equation}
As we have discussed above, the matrix element (\ref{eq:59}) is
generically of ${\cal O}(N_c)$.  However, if $A$ and $A^\prime$ are
pseudoscalars, the baryonic matrix element couples to purely flavor
generators. As a consequence, in the physically relevant case of $B$
and $B^\prime$ being baryonic states with finite flavor (i.e., isospin
and hypercharge of ${\cal O}(N_c^0)$), the matrix element turns out to
be ${\cal O}(N_c^0)$ instead of ${\cal O}(N_c)$. A similar statement
holds for $\omega B\to \omega^\prime B^\prime$ provided $B$ and
$B^\prime$ have finite spin, since $S_i$ is the relevant operator in
this case. For matrix elements of the type $\pi B\to \rho B^\prime$,
the commutation relation
\begin{equation}
[T_{\alpha},G_{\beta i}]= i f_{\alpha\beta\gamma}G_{\gamma i}
\end{equation}
($f_{\alpha\beta\gamma}$ being the flavor structure constants)
indicates that the driving operator is of the type $G_{\alpha i}$,
which is ${\cal O}(N_c)$ even for finite spin-flavor baryons. (Note
that the Casimir operator $T_AT_A$ has a common large value ${\cal
O}(N_c^2)$, to wit, $N_c(N_c+2N_F)(2N_F-1)/(4N_F)$, for all states in
the same irreducible representation ``56''.)  For $\rho B\to
\rho^\prime B^\prime$ the driving operator is
\begin{equation}
[G_{\alpha i},G_{\beta j}]= \frac{i}{4}\delta_{ij}
f_{\alpha\beta\gamma}T_\gamma
+\frac{i}{2}\epsilon_{ijk}
\left(\frac{1}{N_F}\delta_{\alpha\beta}S_k
+d_{\alpha\beta\gamma}G_{\gamma k}\right)
\end{equation}
so generically the matrix element between finite baryons will be large
for $N_F\ge 3$.

As illustration, for two flavors and odd $N_c$, we can consider the
``nucleon'' state with spin and isospin $1/2$
\begin{equation}
|N_{a\sigma}\rangle 
\propto \epsilon^{a_2 a_3}\epsilon^{\sigma_2\sigma_3}\cdots
B^\dagger_{a\sigma,a_2\sigma_2,\ldots} |0\rangle \,, \qquad a,\sigma\in\{1,2\}
\end{equation}
consisting of a single quark carrying the spin and isospin of the
baryon plus $(N_c-1)/2$ pairs of quarks coupled to spin and isospin
zero. An easy computation gives for the ``nucleon'' matrix elements
corresponding to the generators $T_\alpha$, $S_i$ and $G_{\alpha i}$
\begin{equation}
\frac{1}{2}\tau_\alpha \,,
\quad
\frac{1}{2}\sigma_i \,,
\quad
\frac{1}{12}(N_c+2)\tau_\alpha \sigma_i \,,
\end{equation}
respectively, consistently with our previous remarks. Thus the
extension to include vector mesons is indeed essential to activate the
generic large $N_c$ dependences found above.


\subsubsection{Crossed nucleon-pole terms}

\begin{figure}[tbh]
\centerline{\includegraphics[height=3.cm]{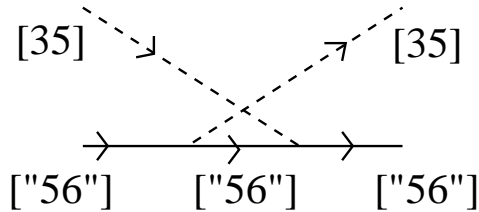}}
\caption{\footnotesize Diagrammatic representation of the crossed 
  nucleon pole-type term.   }\label{fig:fig2}
\end{figure}

The $8_1 \otimes (8_2 \oplus 10_4)$ crossed nucleon pole-type terms,
included among those depicted in Fig.~\ref{fig:fig2}, are believed to
scale as $g_A^2/(M f^2)$ and therefore behave as ${\cal O}(N_c^0)$ in
the large $N_c$ limit~\cite{DJM94}. As we have just seen, the standard
WT term scales as ${\cal O}(N_c^{-1})$, from its $1/f^2$ dependence,
and therefore in the large $N_c$ limit, it is a sub-leading correction
to the crossed nucleon pole-type term. We have shown that this picture
changes when the effects induced by vector mesons and the $N_c$
dependence of the ``56'' irrep are considered, being incorrect within
the ``70'' and ``700'' meson-baryon spaces. Furthermore, one might
wonder whether the $N_c-$behavior of the crossed nucleon pole-type
Hamiltonian, ${\cal H_{\rm CN,SU(6)}^{\rm SU(N_c)}}$, depends on the
SU(6) representation, as it happens in the case of the WT
interaction. To answer such a question, it would be useful to have a
SU(6) symmetric model, for arbitrary $N_c$, for this interaction
term. Because of the $p$-wave nature and the spin dependence of the
$MBB$ coupling, this might not be possible, and at least we have not
been able to come up with a consistent model.  Likely, the spaces that
diagonalize this interaction do not form SU(6) irreps. This is because
in general spin-flavor symmetry is not exact for excited baryons even
in the large $N_c$ \cite{GSS02}. However, phenomenologically for
$N_c=3$, the spin--flavor symmetry breaking term is small and
comparable in magnitude to that of the $1/N_c$
corrections~\cite{GSS02}. Even assuming that ${\cal H_{\rm
CN,SU(6)}^{\rm SU(N_c)}}$ scales as ${\cal O}(N_c^0)$ in the ``70''
irrep space, the WT term provides the large $N_c$ dominant
contribution in this space, since both ${\cal H_{\rm CN,SU(6)}^{\rm
SU(N_c)}}$ and ${\cal H_{\rm WT,SU(6)}^{\rm SU(N_c)}}$ would follow
the same $N_c$ scaling law and the latter one is dominant for $N_c=3$.

\subsubsection{ Resonance Masses and Widths from ${\cal H_{\rm
      WT,SU(6)}^{\rm SU(N_c)}}$}

The resonance equation reads,
\begin{equation}
\beta(s)\Big|_{s=s_R\equiv M^2_R-iM_R\Gamma_R} =  \bar 
\lambda_{\phi_{SU(6)}}, \quad
\phi_{SU(6)}=\makebox{``56'', ``70'', ``700'', ``1134''}, \qquad
M_R>M, \Gamma_R>0
\end{equation} 
There are solutions only for negative eigenvalues, $\bar
\lambda_{\makebox{\footnotesize ``70''}}$, $\bar
\lambda_{\makebox{\footnotesize ``56''}}$ and $\bar
\lambda_{\makebox{\footnotesize ``1134''}}$, and as before, the ``70''
irrep of SU(6) leads to the most attractive $s$-wave meson--baryon
interaction, and it becomes the only non-vanishing WT contribution in the
strict limit $N_c \to \infty$.

\begin{figure}
\vspace{-2cm}
\centerline{\includegraphics[height=25cm]{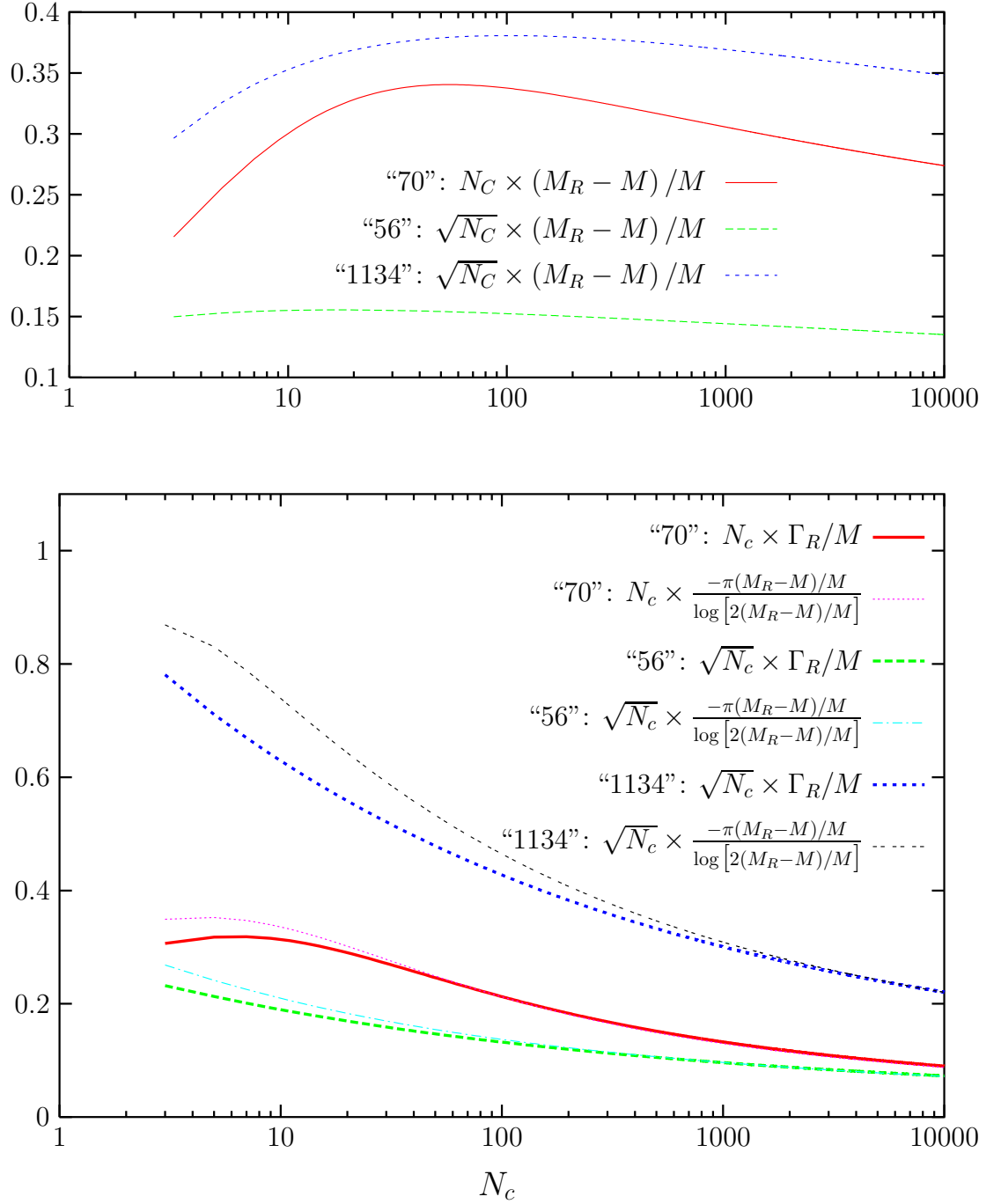}}
\vspace{-5.cm}
\caption{ SU(6) ``70'', ``56'' and ``1134'' resonance masses ($M_R$)
  and widths ($\Gamma_R$) obtained from ${\cal H_{\rm
  WT,SU(6)}^{\rm SU(N_c)}}$, as a function of $N_c$.}
\label{fig:fig3}
\end{figure}

The approximated relations of Eqs.~(\ref{eq:relsu3a})
and~(\ref{eq:relsu3b}), having in mind that  $\bar
\lambda_{\makebox{\footnotesize ``70''}} \sim N_c$, lead to new
scaling relations
\begin{eqnarray}
M_R-M, \, \Gamma_R = {\cal O}(N_c^0)
\label{eq:su6loga}
\end{eqnarray}
for the ``70''--plet. From the above $N_c-$behavior one deduces that
 widths and excitation resonance energies, behave now as order one, as
 predicted by Witten almost 30 years ago. For the ``56'' and
 ``1134''-plets, the scenario has not been modified, and we are still
 in the same situation as in Subsect.~\ref{sec:sec4a}, with widths and
 excitation energies growing as $\sqrt{N_c}$. That is, resonances
 would disappear, since they become wider and heavier as $N_c$
 increases.  The different $N_c$ behaviors exhibited by resonance
 masses and widths, deduced from the WT Lagrangian, in each
 irreducible space can be appreciated in Fig.~\ref{fig:fig3}.

The crossed nucleon pole-type contribution (CNPC) might change this
picture. As discussed at the end of the previous Subsect., we believe
that the CNPC will never be  dominant in the ``70'' irrep space. However,
if we focus on the ``56'' and ``1134'' resonance plets, the WT interaction
could be subleading in the large $N_c$ limit, if the crossed nucleon
pole force would scale as ${\cal O} (N_c^0)$ in those spaces. If this
latter interaction were repulsive, the ``56'' and ``1134'' plets of resonances
would dissappear at sufficiently large values of $N_c$, while if it
were attractive, widths and excitation resonance energies would behave
as order ${\cal O}(N_c^0)$. If the CNPC scales as ${\cal O}
(N_c^{-1})$ or lower in any of those spaces, the corresponding plet of
resonances will either never be formed (if the combined WT
contribution plus CNPC is repulsive) or they will dissappear (become
wider and heavier) at sufficiently large values of $N_c$. For
illustrative purposes in the appendix~\ref{sec:cnpc} we develop a toy
model for the CNPC, somehow unrealistic since it neglects the spin
dependence of the couplings. However, this model shows that is
feasible to have situations in which the CNPC $N_c$ behavior depends on
the particular irrep space and that the WT term provides the large
$N_c$ dominant contribution in the ``70'', ``700'' and ``1134''
irreps  spaces.

Nevertheless, there will be also $d$-wave mixings or new $s$-wave
meson-baryon couplings\footnote{Hamiltonians of the form
(symbolically) $\left((M^\dagger\otimes M)_{35_s}\otimes
(B^\dagger\otimes B)_{35}\right)_{1}$ or $\left((M^\dagger\otimes
M)_{1}\otimes (B^\dagger\otimes B)_{1}\right)_{1}$.} which might also
modify the whole picture.

\section{Concluding Remarks}

\label{sec:sec5}

One of the interesting results of this paper is the Lagrangian in
(\ref{eq:wtnc2}), which accounts for the SU($2N_F$) symmetric version
of the WT interaction, for arbitrary $N_c$ and $N_F$, as well as its
particular case (\ref{eq:lawt}) for $N_c=N_F=3$. As we have noted
above, due to the action of the covariant derivative
(\ref{eq:cov1}--\ref{eq:cov}), it follows that generically (that its,
prior to projection to particular sectors) such extended WT amplitude
scales as ${\cal O}(N_c^0)$, instead of ${\cal O}(N_c^{-1})$,
characteristic of the standard WT SU($N_F$) symmetric amplitude.  Two
factors combine to achieve this result. First, in a large $N_c$ world,
the flavor representation of the lightest baryon depends on $N_c$, and
the standard commutator $[A_3,B_3]$ becomes a covariant derivative,
which acts on each baryon index in turn. Less technically, and using a
graphic quark model picture for the baryon, in the WT interaction, the
meson-meson pair may couple to any of the $N_c$ quarks of the baryon,
allowing a further $N_c$ factor in the amplitude. This mechanism is
also at work in the $p$-wave pseudoscalar-baryon coupling and gives
the standard large $N_c$ scaling $g_A= {\cal O}(N_c)$. However, in the
standard SU($N_F$) case, the pseudoscalar-pseudoscalar amplitude
depends on the flavor generator baryonic matrix element, which is
${\cal O}(N_c)$ for generic baryons but ${\cal O}(N_c^0)$ for the
relevant baryons, namely, those with finite flavor and spin. The
second essential factor is thus the inclusion of vector mesons. They
coupled to spin-flavor generators which are ${\cal O}(N_c)$ even for
baryons with finite flavor and spin.  As a consequence, in the ``70''
and ``700'' SU(6) irreducible spaces, the SU(6) extension of the WT
$s$-wave meson-baryon interaction scales as ${\cal O}(N_c^0)$, instead
of the well known ${\cal O}(N_c^{-1})$ behavior for its SU(3)
counterpart. However, the WT interaction behaves as order ${\cal
O}(N_c^{-1})$ within the ``56'' and ``1134'' meson-baryon spaces.

From constituent quark model considerations, it is accepted that the
excited baryon states that correspond to the first radial and orbital
excitations fit well into respectively a positive parity $56^+$ and
negative parity $70^-$ irreps~\cite{GP76}. From the study carried out
in this work, we confirm the existence of a narrow $70-$plet of
negative parity resonances, which masses depart from the lowest--lying
56 multiplet baryon mass by the typical amount of a meson mass. The
non existence, in the large $N_c$ limit, of negative parity $56^-$
resonances can be understood if the crossed nucleon pole force is
repulsive or if it is attractive, it should decrease at least as
${\cal O}(N_c^{-1})$ in this irrep. Thus, one of the two
$\frac{1}{2}^-$ SU(3) octets of $s$-wave baryon resonances found in
Ref.~\cite{Ga03} would dissappear in the large $N_c$ limit, since
there exists only one $8_2$ multiplet included into the 70
representation of SU(6). However, the SU(3) singlet spin-parity
$\frac12^-$ resonance will become presumably narrow, in the large
$N_c$ limit, thanks to its 70 component.

On the other hand, the WT interaction predicts for the 1134-plet that
both excitation energies and widths grow with an approximate
$\sqrt{N_c}$ rate.  This presumably implies that these states do not
appear in the large $N_c-$QCD spectrum, which most likely reflects the
existence of exotic, f.i. $qqqq\bar q$, components\footnote{A similar
analysis has been carried out in Ref.~\cite{Pe04} in the the
meson--meson context. There, it was found that while the $\rho$ and
$K^*$ meson widths have the $q \bar q$ expected behavior (${\cal O}
(1/N_c)$), the $s$-wave $\sigma$ and $\kappa$ poles show a totally
different behavior, since their widths grow with $N_c$, in conflict
with a $q \bar q$ interpretation, and leaving room for sizeable
tetra-quark or glue ball components.} for $N_c=3$. Note that exotic
components are certainly included in the SU(3) antidecuplet belonging
to the 1134 SU(6) representation.

Finally, as we have noted previously, in the present approach the
power-like $1/N_c$ expansion comes with subleading logarithmic
corrections (see for instance Eqs.~(\ref{eq:loga}-\ref{eq:logb})),
which are believed to be spurious. It remains to be studied in deep,
how the logarithmic corrections depend on details of the RS
prescription, the baryon wavefunction renormalization, etc. This
subject is worth studying and clearly it would be highly desirable to
consider this issue for future research, however, is beyond the scope
of the present work.

\begin{acknowledgments}
We warmly thank to Dr. E. Ruiz Arriola  for useful
discussions. This research was supported by DGI and FEDER funds, under contract
FIS2005-00810, by the Junta de Andaluc\'\i a (FQM-225) and the EU network
EURIDICE, under contract HPRN-CT-2002-00311. It is also part of the EU
integrated infrastructure initiative Hadron Physics Project under
contract number RII3-CT-2004-506078.

\end{acknowledgments}

\appendix 
\section{Some details on $\chi-$BS(3)}
\label{sec:app}

For a given isospin--strangeness sector (for
simplicity we will omit the $IS$ upper indices), the element in the
position $aa ~~(a=1,..N^{IS})$ of the diagonal matrix loop function
$J(\sqrt{s})$ reads~\cite{nrg}
\begin{eqnarray}
J_a(\sqrt{s}) &=&{ (\sqrt{s}+ M_a)^2- m_a^2  \over 2 \sqrt{s}}
 {\cal J}_a(s)\label{eq:loopf} 
\end{eqnarray}
where $M_a (m_a)$ is the baryon (meson) mass in the channel $a$ and 
\begin{eqnarray}
{\cal J}_a(s)  &=&  i \int { d^4 q \over (2\pi)^4 } {1\over
q^2 - m_a^2}{1\over (P-q)^2 -  M_a^2 }= \bar {\cal J}_a( s) 
+ {\cal J}_a ( s= (m_a +  M_a)^2) \label{eq:defj0}
\end{eqnarray}
with $P^2=s$, ${\cal J}_a  (\ s= (m_a+ M_a)^2)$ a divergent quantity and the
finite function $\bar {\cal J}_a( s)  $  given
by
\begin{eqnarray}
\bar {\cal J}_a( s)  &=& {1\over (4\pi)^2} \left\{ \left[ {M_a^2
-m_a^2 \over s} -{M_a -m_a \over M_a + m_a } \right] {\rm ln} {M_a
\over m_a } +  L_a( s) \right\}
\end{eqnarray} 
and for real $s$ and above threshold, $s>(m_a+M_a)^2$, we have
\begin{eqnarray} 
L_a(  s) &\equiv& L_a( s+ i \epsilon ) =
{\lambda^{1/2}(s,m^2_a,M^2_a) \over s} \left\{ \log \left[{ 1+ \sqrt{s-s_+
\over s-s_- } \over 1-\sqrt{s-s_+ \over s-s_- } } \right] -i\pi
\right\}
\end{eqnarray}  
where $\lambda(x,y,z) = x^2+y^2+z^2-2xy-2xz-2yz$, the pseudothreshold
and threshold variables are $s_{\mp} = (M_a \mp m_a)^2$ respectively,
and the logarithm is taken to be real. Note that $L_a(s_+) =0$.  The
definition of the $L_a(  s)$ in the whole complex plane and the
definition of its different Riemann sheets can be found in
Ref.~\cite{nrg}.

\section{Restriction of ${\cal L}_{\rm SU(6)}$ to the $8_1 \otimes
  8_2$ sector}
\label{sec:restr}

In this appendix we check that the restriction of the SU(6) Lagrangian
of Eq.~(\ref{eq:lawt}) to the $8_1 \otimes 8_2$ sector reproduces that
given in Eq.~(\ref{eq:su3su2}), which provides the standard WT
amplitudes of Eq.~(\ref{eq:lowest}). We will do it for three flavors,
though the extension to $N_F$ flavors is straightforward. We will
start studying the meson part of the Lagrangian.

The operator $M^i_j$ (see Eq.~(\ref{eq:defm})) is
essentially the annihilation part of the meson matrix
$[\Phi_6]^i_j$. The projection $\left( \equiv (\Phi_6)_3\right)$ of
$\Phi_6$ to the $8_1$ octet is
\begin{equation}
[(\Phi_6)_3]^a_b = \frac{1}{\sqrt{2}} \sum_{\sigma=1,2}
[\Phi_6]^{a\sigma}_{b\sigma} = \frac{1}{\sqrt 2}[{\rm Tr}_{\rm
SU(2)}(\Phi_6)]^a_b, \quad a,b=1,2,3
\end{equation}
that is, in the above equation $a,b$ account for the quark and
antiquark flavors, while for quark (antiquark), $\sigma=1$ corresponds
to $S_z= 1/2\, (-1/2)$ and $\sigma=2$ corresponds to $S_z=
-1/2\,(1/2)$. Thus, for instance $[(\Phi_6)_3]^1_1 =
\left([\Phi_6]^1_1+ [\Phi_6]^4_4\right)/\sqrt{2}= \pi^0/\sqrt{2}
+\eta/\sqrt{6}$, as one can deduce from $\left (M^1_1
+M^4_4\right)/\sqrt{2}$. Reciprocally, the contribution $\left( \equiv
(\Phi_3)_6\right)$ of the $8_1$ octet to $\Phi_6$ is
\begin{equation}
[(\Phi_3)_6]^{a\sigma}_{b\sigma^\prime} = \frac{1}{\sqrt{2}} [\Phi_3]^a_b
 \times \delta^\sigma_{\sigma^\prime}
\end{equation}
and thus we have $(\Phi_3)_6= \left (\Phi_3 \otimes 1_{\rm
SU(2)}\right)/\sqrt{2}$. We see that the consistency relations 
$((\Phi_3)_6)_3 = \Phi_3$ and Tr$\left [(\Phi_3^\prime)_6\cdot
  (\Phi_3)_6\right ] =
{\rm Tr}[\Phi^\prime_3 \cdot \Phi_3]$ are trivially satisfied. On the
other hand, when $\Phi_6=(\Phi_3)_6$ we must require $U_6= U_3\otimes 1_{\rm
SU(2)}$, then 
\begin{equation}
f_6 = \frac{f}{\sqrt{2}} \label{eq:f6}
\end{equation}
 Besides, it is also satisfied that
\begin{equation}
(A^\mu_3)_6= A_3^\mu \otimes 1_{\rm SU(2)}
\end{equation}

For SU(3) and $N_c=3$, the $B_3$ field is normalized such that
\begin{equation}
{\rm Tr} (B^\dagger_3 B_3) = \sum_{\lambda} b_\lambda^\dagger
b_\lambda =  p^\dagger p+ n^\dagger n+ \Lambda^\dagger \Lambda+
\Sigma^{0\dagger}\Sigma^0+ \Sigma^{+\dagger}\Sigma^++
\Sigma^{-\dagger}\Sigma^-+ \Xi^{0\dagger}\Xi^0 + \Xi^{-\dagger}\Xi^- 
\end{equation}
with $b_\lambda$ a $8_2$ baryon field. The  normalization above
is consistent with that adopted for the SU(6) fields in
Eq.~(\ref{eq:su6norm}).  On the other hand, we will denote the $B_3$
field indices as ${[B_3]^a_.}_b^{\phantom{a}\sigma}, a,b=1,2,3$ and
$\sigma = \pm $ accounts for the spin third component of the baryon
($\pm 1/2)$, while for $B_6$ we will use ${\cal B}^{ijk},
i,j,k=1,\ldots 6 $. The projection $\left( \equiv (B_6)_3\right)$ of
$B_6$ to the $8_2$ octet is (up to a global sign)
\begin{equation}
{[(B_6)_3]^a_.}_b^{\phantom{a}\sigma} = \frac{1}{2\sqrt{3}} {\cal
  B}^{a\sigma^{\phantom{\prime}} c\sigma^\prime d\sigma^{\prime\prime}}
  \epsilon_{bcd}\,\epsilon_{\sigma^\prime\sigma^{\prime\prime}}, \quad
  a,b,c,d= 1,2,3, \quad
  \sigma,\sigma^\prime,\sigma^{\prime\prime}=\pm.
\end{equation}
with $\epsilon_{++}=\epsilon{--}=0$ and
$\epsilon_{+-}=-\epsilon_{-+}=1$. This definition  satisfies
${[(B_6)_3]^a_.}_a^{\phantom{a}\sigma}=0$, as required to ensure that
those states belong to the octet irrep of SU(3), and the normalization
can be easily tested by checking, for instance, that
\begin{equation}
\underwick{1}{[(<1B_6)_3^\dagger]^{1+}_3\,[(>1B_6)_3]^{1+}_3} = 1, \qquad
\underwick{1}{[(<1B_6)_3^\dagger]^{1+}_1\,[(>1B_6)_3]^{1+}_1} = 2/3 
\end{equation}
which correctly accounts for the normalization of a diagonal,
$p(S_z=+1/2)$, and of a non-diagonal, $\left\{\sqrt{3}\Sigma^0(S_z=+1/2)+
\Lambda(S_z=+1/2)\right\}/\sqrt{6}$, elements of the $B_3$
matrix. Reciprocally, the contribution $\left( \equiv
(B_3)_6\right)$ of the $8_2$ octet to $B_6$ is
\begin{equation}
(B_3)_6 \to ({\cal
    B}_3)_6^{a\sigma^{\phantom{\prime}}
    b\sigma^{\prime}c\sigma^{\prime\prime}} = \frac{1}{\sqrt{3}} \left
    ( {[B_3]^a_.}_d^{\phantom{a}\sigma}
    \epsilon^{dbc}\epsilon^{\sigma^\prime \sigma^{\prime\prime}} + 
{[B_3]^b_.}_d^{\phantom{a}\sigma^\prime}
    \epsilon^{dac}\epsilon^{\sigma \sigma^{\prime\prime}} + 
{[B_3]^c_.}_d^{\phantom{a}\sigma^{\prime\prime}}
    \epsilon^{dab}\epsilon^{\sigma \sigma^{\prime}}
 \right ) 
\end{equation}
With these definitions it is straightforward to prove the consistency
relations  $((B_3)_6)_3 = B_3$, and\footnote{To deduce
  Eq.~(\ref{eq:a3a6}) we have made use of 
\begin{eqnarray}
\Big[(A^\mu_3)_6 * (B_3)_6\Big]^{a\sigma^{\phantom{\prime}}
  b\sigma^\prime c\sigma^{\prime\prime}} &=& [(A^\mu_3)_6]^{i_1}_k
  [({\cal B}_3)_6]^{ki_2i_3} + [(A^\mu_3)_6]^{i_2}_k [({\cal
  B}_3)_6]^{i_1ki_3}+[(A^\mu_3)_6]^{i_3}_k [({\cal B}_3)_6]^{i_1i_2k}
  \nonumber \\ &=& \frac{1}{\sqrt{3}} \left( [A^\mu_3]^a_e
  \,{[B_3]^e_.}_d^{\phantom{a}\sigma}\epsilon^{dbc}\epsilon^{\sigma^\prime\sigma^{\prime\prime}}
  + [A^\mu_3]^a_e
  \,{[B_3]^b_.}_d^{\phantom{a}\sigma^\prime}\epsilon^{dec}\epsilon^{\sigma\sigma^{\prime\prime}}
+ [A^\mu_3]^a_e
  \,{[B_3]^c_.}_d^{\phantom{a}\sigma^{\prime\prime}}\epsilon^{deb}\epsilon^{\sigma\sigma^{\prime}}
  +  \begin{array}{ccc}a &\leftrightarrow & b \\ \sigma
  &\leftrightarrow & \sigma^\prime\end{array} +  \begin{array}{ccc}a &\leftrightarrow & c \\ \sigma
  &\leftrightarrow & \sigma^{\prime\prime}\end{array} \right)
  \nonumber\\
&=& \frac{1}{\sqrt{3}} \left( [A^\mu_3]^a_e
  \,{[B_3]^e_.}_d^{\phantom{a}\sigma}\epsilon^{dbc}\epsilon^{\sigma^\prime\sigma^{\prime\prime}}
  + 2 [A^\mu_3]^a_e
  \,{[B_3]^b_.}_d^{\phantom{a}\sigma^\prime}\epsilon^{dec}\epsilon^{\sigma\sigma^{\prime\prime}}
 +  \begin{array}{ccc}a &\leftrightarrow & b \\ \sigma
  &\leftrightarrow & \sigma^\prime\end{array} +  \begin{array}{ccc}a &\leftrightarrow & c \\ \sigma
  &\leftrightarrow & \sigma^{\prime\prime}\end{array} \right)\nonumber\\
&=& \frac{1}{\sqrt{3}} \left( [A^\mu_3]^a_e
  \,{[B_3]^e_.}_d^{\phantom{a}\sigma}\epsilon^{dbc}\epsilon^{\sigma^\prime\sigma^{\prime\prime}}
  + 2 [A^\mu_3]^b_e
  \,{[B_3]^a_.}_d^{\phantom{a}\sigma}\epsilon^{dec}\epsilon^{\sigma^\prime\sigma^{\prime\prime}}
 +  \begin{array}{ccc}a &\leftrightarrow & b \\ \sigma
  &\leftrightarrow & \sigma^\prime\end{array} +  \begin{array}{ccc}a &\leftrightarrow & c \\ \sigma
  &\leftrightarrow & \sigma^{\prime\prime}\end{array} \right)\label{eq:con1}
\end{eqnarray}
where we associate to the SU(6) indices $i_1,i_2$ and $i_3$, the
SU(3)$\times$ SU(2) $a\sigma^{\phantom{\prime}},b\sigma^{\prime}$ and
$c\sigma^{\prime\prime} $ ones, respectively. Because of the
antisymmetric tensor $\epsilon^{\sigma^\prime\sigma^{\prime\prime}}$, 
and the underlying symmetry under the
interchange $(i_1 \leftrightarrow i_2$), one should only keep 
the antisymmetric contribution, when the
SU(3) indices $b$ and $c$ are interchanged, of the second term in
Eq.~(\ref{eq:con1}), thus we find
\begin{eqnarray}
\Big[(A^\mu_3)_6 * (B_3)_6\Big]^{a\sigma^{\phantom{\prime}}
  b\sigma^\prime c\sigma^{\prime\prime}} &=& \frac{1}{\sqrt{3}} \left(
  [A^\mu_3]^a_e
  \,{[B_3]^e_.}_d^{\phantom{a}\sigma}\epsilon^{dbc}\epsilon^{\sigma^\prime\sigma^{\prime\prime}}
  + [A^\mu_3]^b_e
  \,{[B_3]^a_.}_d^{\phantom{a}\sigma}\epsilon^{dec}\epsilon^{\sigma^\prime\sigma^{\prime\prime}}
  -[A^\mu_3]^c_e
  \,{[B_3]^a_.}_d^{\phantom{a}\sigma}\epsilon^{deb}\epsilon^{\sigma^\prime\sigma^{\prime\prime}}
  + \begin{array}{ccc}a &\leftrightarrow & b \\ \sigma
  &\leftrightarrow & \sigma^\prime\end{array} + \begin{array}{ccc}a
  &\leftrightarrow & c \\ \sigma &\leftrightarrow &
  \sigma^{\prime\prime}\end{array} \right) \nonumber \\ &=&
  \frac{1}{\sqrt{3}} \left ( {[A_3^\mu,B_3]^a_.}_d^{\phantom{a}\sigma}
  \epsilon^{dbc}\epsilon^{\sigma^\prime\sigma^{\prime\prime}} +
  \begin{array}{ccc}a &\leftrightarrow & b \\ \sigma &\leftrightarrow
  & \sigma^\prime\end{array} + \begin{array}{ccc}a &\leftrightarrow &
  c \\ \sigma &\leftrightarrow & \sigma^{\prime\prime}\end{array}
  \right)
\end{eqnarray}
thanks to the relation
\begin{equation}
[A_3^\mu]^b_e~\epsilon^{dec} - [A_3^\mu]^c_e~\epsilon^{deb} =
- [A_3^\mu]^d_e~\epsilon^{ebc} 
\end{equation}
}
\begin{eqnarray}
(B_3^\dagger)_6\cdot(B_3)_6 &=&
  {[B_3^\dagger]^a_.}_{b\sigma} {[B_3]^b_.}_a^{\phantom{a}\sigma} =
  {\rm Tr}(B^\dagger_3 B_3), \label{eq:tr}\\
\Big[(A^\mu_3)_6 * (B_3)_6\Big]^{a\sigma^{\phantom{\prime}}
  b\sigma^\prime c\sigma^{\prime\prime}} &=& \frac{1}{\sqrt{3}} \left
  ( {[A_3^\mu,B_3]^a_.}_d^{\phantom{a}\sigma}
  \epsilon^{dbc}\epsilon^{\sigma^\prime\sigma^{\prime\prime}} +
{[A_3^\mu,B_3]^b_.}_d^{\phantom{a}\sigma^\prime}
  \epsilon^{dac}\epsilon^{\sigma\sigma^{\prime\prime}}+
{[A_3^\mu,B_3]^c_.}_d^{\phantom{a}\sigma^{\prime\prime}}
  \epsilon^{dab}\epsilon^{\sigma\sigma^{\prime}}  \right) \nonumber \\
&=& (A_3^\mu * B_3)_6^{a\sigma^{\phantom{\prime}}
  b\sigma^\prime c\sigma^{\prime\prime}}, \qquad {\rm with}~~
A_3^\mu  * B_3 = [A_3^\mu, B_3]  \label{eq:a3a6}
\end{eqnarray}
where $B_6^\dagger  \cdot B_6 \equiv
\frac{1}{3!} {\cal B}^\dagger_{i_1i_2i_3} {\cal
B}^{i_1i_2i_3}$.  From the above equations, it follows 
\begin{equation}
(B_3^\dagger)_6\cdot\left(  (A^\mu_3)_6 * (B_3)_6 \right ) =
  (B_3^\dagger)_6\cdot(A_3^\mu * B_3)_6 = {\rm Tr}(B^\dagger_3\cdot
  (A^\mu_3* B_3)) = {\rm Tr}(B^\dagger_3 [A^\mu_3, B_3]) 
\end{equation}
This latter equation together Eq.~(\ref{eq:tr}), shows that the
restriction of the Lagrangian in Eq.~(\ref{eq:lawt}) to the $8_1
\otimes 8_2$ sector reproduces that given in Eq.~(\ref{eq:su3su2}).

\section{Eigenvalues of ${\cal G_{\rm WT,SU(6)}^{\rm SU(N_c)}}$}

\label{sec:gwt}

Since the 35-meson--``56''-baryon hamiltonian ${\cal H_{\rm
WT,SU(6)}^{\rm SU(N_c)}}$ (or equivalently ${\cal G_{\rm WT,SU(6)}^{\rm
    SU(N_c)}}$)  is a SU(6) scalar, its eigenvectors follow
from the SU(6) reduction given in Eq.~(\ref{eq:su6_nc}).

A meson-baryon state belonging to the ``56'' representation is of
the form
\begin{equation}
\sum_{P\in S(N_c)}M^{\dagger m}_{P(i_1)} B^\dagger_{m P(i_2)\ldots
  P(i_{N_c})}|0\rangle 
\end{equation}
where $|0\rangle $ is the ground state in the Fock space (state
containing zero hadrons). All of these states are eigenstates of
${\cal G_{\rm WT,SU(6)}^{\rm SU(N_c)}}$, with eigenvalue proportional
to $\bar\lambda_{\makebox{\rm \footnotesize ``56''}}$, which might
depend on $N_c$.  One of these $\left (\begin{array}{c}N_c+5\cr
5\end{array}\right)$ states is
\begin{equation}
|1\rangle = M^{\dagger m}_1 B^\dagger_{m 11\ldots 1} |0\rangle
\end{equation}
One finds, 
\begin{eqnarray}
{\cal G_{\rm  WT,SU(6)}^{\rm SU(N_c)}} |1\rangle &=&
\frac{2}{(N_c-1)!}
\left(
M^{\dagger j}_k
 \underwick{1}{<1M^i_j >1M^{\dagger m}_1}
 -M^{\dagger i}_j
 \underwick{1}{<1M^j_k >1M^{\dagger m}_1}
\right)
B^\dagger_{ii_2\ldots i_{N_c}}
\underwick{1}{<1B^{ki_2\ldots i_{N_c}}>1
B^\dagger_{m1\ldots
1}} |0\rangle 
\nonumber \\
&=& \left (-4N_F \right) |1\rangle
\end{eqnarray}
where we have made use of the Wick's contractions given in
Eq.~(\ref{eq:wick}). 

Therefore $\bar\lambda_{\makebox{\rm
\footnotesize ``56''}} = -4N_F$. Analogously and using the states
\begin{eqnarray} 
|2\rangle &=&  \left (
 M^{\dagger i}_2 B^\dagger_{i11\ldots 1} -M^{\dagger i}_1
 B^\dagger_{i21\ldots 1} \right )|0\rangle \nonumber \\
|3\rangle &=&  M^{\dagger 2}_1
 B^\dagger_{111\ldots 1}|0\rangle \nonumber \\
|4 \rangle &=& \left (M^{\dagger 3}_2
 B^\dagger_{111\ldots 1} - M^{\dagger 3}_1
 B^\dagger_{211\ldots 1} \right )|0\rangle
\end{eqnarray}
for the ``70'', ``700'' and ``1134'' eigenspaces respectively, we
find the eigenvalues of ${\cal G_{\rm  WT,SU(6)}^{\rm SU(N_c)}}$ in
these spaces. Thus, we finally conclude
\begin{equation}
\left [\bar\lambda_{\makebox{\rm
\footnotesize ``56''}},\, \bar\lambda_{\makebox{\rm \footnotesize
      ``70''}},\, \bar\lambda_{\makebox{\rm \footnotesize ``700''}},\, \bar\lambda_{\makebox{\rm
\footnotesize ``1134''}} \right ]   = \left  [  -4N_F,\, -2(N_c+2N_F) ,\, 2N_c,\, 
 - 2 \right ]
\end{equation} 

\section{A toy model for the $N_c$ dependence of the  ${\cal H}_{CN}$
 SU(6) eigenvalues}

\label{sec:cnpc}

For illustrative purposes, in what follows we develop a simple model,
where we ignore the spin dependence of the couplings. Symbolically,
the crossed nucleon pole-type hamiltonian, ${\cal H}_{CN}$, might take
the form (see diagram of Fig.~\ref{fig:fig2})
\begin{equation}
{\cal H}_{CN}= \left((M^\dagger\otimes B)_{
\makebox{\footnotesize  ``56''}} \otimes 
(M\otimes B^\dagger)_{ \makebox{\footnotesize ``56}^*
  \makebox{\footnotesize ''}}\right)_1
\end{equation}
Since
\begin{equation}
(M^\dagger\otimes B)_{\makebox{\footnotesize ``56''}}^{i_1i_2\ldots i_{N_c}} =
  \frac{1}{N_c!} \sum_{P\in S_{N_c}} B^{jP(i_2)P(i_3)\ldots P(i_{N_c})}
  M^{\dagger P(i_1)}_j 
\end{equation}
the group structure, ${\cal G_{\rm CN,SU(6)}^{\rm SU(N_c)}}$, of the
    crossed nucleon pole-type hamiltonian, up to constant factors and
    within this simplified model, would take the form
\begin{eqnarray}
{\cal G_{\rm CN,SU(6)}^{\rm SU(N_c)}}&=& \frac{1}{(N_c-1)!} \frac{1}{N_c!} 
 :  \sum_{P\in S_{N_c}} B^{jP(i_2)P(i_3)\ldots P(i_{N_c})}
  M^{\dagger P(i_1)}_j  B^\dagger_{li_2i_3\ldots i_{N_c}}M^l_{i_1}:
  \nonumber \\
&=& :\frac{1}{N_c!} B^{ji_2i_3\ldots i_{N_c}}
  M^{\dagger k}_j  B^\dagger_{li_2i_3\ldots i_{N_c}}M^l_k +
  \frac{N_c-1}{N_c!}   B^{jki_3\ldots i_{N_c}}
  M^{\dagger r}_j  B^\dagger_{lri_3\ldots i_{N_c}}M^l_k: \label{eq:modelcn}
\end{eqnarray}
From the eigenvalues of ${\cal G_{\rm CN,SU(6)}^{\rm SU(N_c)}}$, we
obtain the following proportionality relations for those of ${\cal
H_{\rm CN,SU(6)}^{\rm SU(N_c)}}$ in this model\footnote{The values
quoted within brackets are the eigenvalues of ${\cal G_{\rm CN,SU(6)}^{\rm
SU(N_c)}}$.}
\begin{equation}
\left[\lambda^{\rm CN}_{\makebox{\rm
\footnotesize ``56''}},\, \lambda^{\rm CN}_{\makebox{\rm \footnotesize
      ``70''}},\, \lambda^{\rm CN}_{\makebox{\rm \footnotesize ``700''}},\, 
\lambda^{\rm CN}_{\makebox{\rm
\footnotesize ``1134''}} \right]   \propto  
\left[ 
N_c +2N_F -1 -\frac{N_c}{2N_F} -\frac{2N_F}{N_c}
,\,
-1-\frac{2N_F}{N_c} ,\, 
1 ,\, 
- \frac{1}{N_c}\right] \,.
\end{equation} 
We see that in the large $N_c$ limit both, $\lambda^{\rm
CN}_{\makebox{\rm \footnotesize ``70''}}/\lambda^{\rm
CN}_{\makebox{\rm \footnotesize ``56''}}$ and $\lambda^{\rm
CN}_{\makebox{\rm \footnotesize ``700''}}/\lambda^{\rm
CN}_{\makebox{\rm \footnotesize ``56''}}$, behave as ${\cal
O}(1/N_c)$, while $\lambda^{\rm CN}_{\makebox{\rm \footnotesize
``1134''}}/\lambda^{\rm CN}_{\makebox{\rm \footnotesize ``56''}}$ is
suppressed by $N_c^{-2}$. Thus in this model, we find that the crossed
nucleon pole-type contribution depends on the SU(6) representation,
being the ``56'' the dominant one. Assuming that $\lambda^{\rm
CN}_{\makebox{\rm \footnotesize ``56''}}$ scales as ${\cal O}(N_c)$,
we find that the WT term provides the large $N_c$ dominant
contribution in the ``70'', ``700'' and ``1134'' irreducible
representation spaces. Indeed within this toy model, the crossed
nucleon pole-type contributions in those spaces are suppressed by
$1/N_c$ with respect the WT ones.



\begin{thebibliography}{99}

\bibitem{Ho74} G.'t Hooft, Nucl. Phys. {\bf B72} (1974) 461; {\it
ibidem} Nucl. Phys. {\bf B75} (1974) 461.

\bibitem{Wi79} E. Witten, Nucl. Phys. {\bf B160} (1979) 57.

\bibitem{Ma99} 
A.V. Manohar, {\it Large $N_c$ QCD} in Probing the
Standard Model of Particle Interactions, Les Houches LXVIII (1997),
Elsevier, Amsterdam, 1999, Vol. II., p. 1091, hep-ph/9802419. 

\bibitem{Ga03} C. Garc\'\i a-Recio, M.F.M Lutz and J. Nieves, 
Phys. Lett. {\bf B582} (2004) 49.

\bibitem{KL03} E.E. Kolomeitsev and M.F.M. Lutz, Phys. Lett. {\bf
  B585} (2004) 243.

\bibitem{LK02} M. F. M. Lutz and E. E. Kolomeitsev, Nucl. Phys. 
{\bf A700} (2002) 193.

\bibitem{Lu03} M.F.M Lutz, C. Garc\'\i a-Recio,
E.E. Kolomeitsev and J. Nieves, Nucl. Phys. {\bf A754} (2005) 212c.

\bibitem{LK04} M.F.M. Lutz and E.E. Kolomeitsev,  Nucl.Phys. {\bf
A730} (2004) 392.

\bibitem{LK04charm}
M.F.M. Lutz and E.E. Kolomeitsev, Nucl. Phys.  {\bf A730} (2004) 110;
J. Hofmann and M.F.M. Lutz, Nucl. Phys.  {\bf A763} (2005) 90.

\bibitem{DJM94} R. Dashen, E. Jenkins and A.V. Manohar,
  Phys. Rev. {\bf D49} (1994) 4713.

\bibitem{GNS05} C. Garc\'\i a--Recio, J. Nieves and L.L. Salcedo, 
hep-ph/0505233.

\bibitem{Jenkins:1990jv}
  E.~Jenkins and A.~V.~Manohar,
  Phys.\ Lett.\ B {\bf 255} (1991) 558.

\bibitem{Ellis:1997kc}
  P.~J.~Ellis and H.~B.~Tang,
  Phys.\ Rev.\ C {\bf 57} (1998) 3356
  [arXiv:hep-ph/9709354].

\bibitem{Becher:1999he}
  T.~Becher and H.~Leutwyler,
  Eur.\ Phys.\ J.\ C {\bf 9} (1999) 643
  [arXiv:hep-ph/9901384].

\bibitem{Wein-Tomo} S. Weinberg, Phys. Rev. Lett. {\bf 17} (1966) 616;
Y. Tomozawa, Nuov. Cim. {\bf A46} (1966) 707.

\bibitem{weise} N. Kaiser, P.B. Siegel and W. Weise, Nucl. Phys. {\bf
A594} (1995) 325; {\it ibidem} Phys. Lett. {\bf B362} (1995) 23.

\bibitem{oset} E. Oset and A. Ramos, Nucl. Phys. {\bf A635} (1998) 99;
E. Oset, A. Ramos and C. Bennhold, Phys. Lett. {\bf B527} (2002) 99;
A. Ramos, E. Oset and C. Bennhold, Phys. Rev. Lett. {\bf 89} (2002)
252001; T. Inoue, E. Oset and M.J. Vicente-Vacas, Phys. Rev. {\bf C65}
(2002) 035204.

\bibitem{OM} J. Oller and U. Mei\ss ner, Phys. Lett. {\bf 
B500} (2001) 263.

\bibitem{nrg} J. Nieves and E. Ruiz Arriola, Phys. Rev. {\bf D64},
(2001) 116008; C. Garc\'\i a-Recio, J. Nieves, E. Ruiz Arriola and
M. J. Vicente-Vacas, Phys. Rev.  {\bf D67} (2003) 076009.

\bibitem{Pich95} A. Pich, Rep. Prog. Phys. {\bf 58} (1995) 563.

\bibitem{EJmeson} J. Nieves and E. Ruiz Arriola,  Phys. Lett.
{\bf B455} 30 (1999);  Nucl. Phys. {\bf A679} 57 (2000).

\bibitem{Ji03}
D. Jido, J.A. Oller, E. Oset, A. Ramos and U.G. Meissner,
Nucl. Phys.  {\bf A725} (2003) 181.

\bibitem{ORM05} V.K. Magas, E. Oset and  A. Ramos,
  Phys. Rev. Lett. {\bf 95} (2005) 052301.

\bibitem{PY98} D. Pirjol and T.-M. Yan, Phys. Rev. {\bf D57} (1998) 1449.


\bibitem{Co04} T.D. Cohen, D. C. Dakin, A. Nellore and R.F. Lebed,
  Phys. Rev. {\bf D69} (2004) 056001.

\bibitem{GSS05} J.L. Goity, C.L. Schat and N.N. Scoccola,
  Phys. Rev. {\bf D71} (2005) 034016; J.L. Goity,
  Phys. Atom. Nucl. {\bf 68} (2005) 624.

\bibitem{Ope} C.E. Carlson, C.D. Carone, J.L. Goity and R.F. Lebed,
  Phys. Lett. {\bf B438} (1998) 327; C.E. Carlson, C.D. Carone,
  J.L. Goity and R.F. Lebed, Phys. Rev. {\bf D59} (1999) 114008;
  C.D. Carone, H. Georgi, L. Kaplan, and D. Morin, Phys. Rev. {\bf
  D50} (1994) 5793; J.L. Goity, Phys. Lett. {\bf B414} (1997)
  140. C.E. Carlson and C.D. Carone, Phys. Lett. {\bf B484} (2000)
  260; J.L. Goity, C.L. Schat and N.N. Scoccola, Phys. {\bf B564}
  (2003) 83.

\bibitem{GSS02}  C.L. Schat, J.L. Goity and N.N. Scoccola,
  Phys. Rev. Lett. {\bf 88} (2002) 102002; J.L. Goity, C.L. Schat and
  N.N. Scoccola Phys. Rev. {\bf D66} (2002) 114014.


\bibitem{Solitons} A. Hayashi, G. Eckart, G. Holzwarth, H. Walliser,
  Phys. Lett. {\bf B147} (1984) 5.; M.P. Mattis and M. Karliner,
  Phys. Rev.  {\bf D31} (1985) 2833; M.P. Mattis and M.E. Peskin,
  Phys. Rev.  {\bf D32} (1985) 58; M.P. Mattis,
  Phys. Rev.  Lett. {\bf 56} (1986) 1103; M.P. Mattis and M. Mukerjee,
  Phys. Rev.  Lett. {\bf 61} (1988) 1344; M.P. Mattis,
  Phys. Rev.  {\bf D39} (1989) R994; M.P. Mattis,
  Phys. Rev.  Lett. {\bf 63} (1989) 1455.

\bibitem{CL03} T.D. Cohen and R.F. Lebed, Phys. Rev. {\bf D68} (2003) 056003.



\bibitem{CP76} D.G. Caldi and H.Pagels, Phys. Rev. {\bf D14} (1976)
  809;  Phys. Rev. {\bf D15} (1977) 2668.

\bibitem{CG-su3} J.J. de Swart, Rev. Mod. Phys. {\bf 35} (1963) 916.

\bibitem{CG-su6} J.C. Carter, J.J. Coyne and S. Meshkov,
  Phys. Rev. Lett. {\bf 14} (1965) 523; {\it erratum}
  Phys. Rev. Lett. {\bf 14} (1965) 850.

\bibitem{PIN} M. Mojzis, Eur. Phys. Jour. {\bf C2} (1998) 181;
  N. Fettes and U.-G. Mei\ss ner, Nucl. Phys. {\bf A676} (2000) 311;
  A. G\'omez-Nicola, J. Nieves, J.R. Pel\'aez and E. Ruiz-Arriola,
  Phys. Rev. {\bf D69} (2004) 076007; Phys. Lett. {\bf B486} (2000)
  77.

\bibitem{CL04} T.D. Cohen and R.F. Lebed, Phys. Rev. {\bf D70} (2004) 096015.

\bibitem{CM67} S. Coleman and J. Mandula, Phys. Rev. {\bf 159} (1967)
  1251.

\bibitem{GP76} W.M. Gibson and B.R. Pollard, {\it Symmetry Principles
  in Elementary Particle Physics}, Cambrigdge University Press, 1976.

\bibitem{Pe04} J.R. Pelaez, Phys. Rev. Lett. {\bf 92} (2004) 102001. 

\end{thebibliography}
\end{document}